\documentclass{aastex63}
\usepackage{comment}
\usepackage{array,multirow}

\newcommand{\TextUnderscore}{\rule{.5em}{.3pt}}

\shorttitle{Structural parameters and radial velocity dispersion profile of NGC 6569}
\shortauthors{}
\graphicspath{{./}{figures/}}

\begin{document}

\title{Internal kinematics and structure of the bulge globular cluster
  NGC 6569\footnote{Based on observations collected at the European
  Southern Observatory, Cerro Paranal (Chile), in the context of the
  ESO-VLT Multi-Instrument Kinematic Survey (MIKiS survey) under Large
  Programmes 106.21N5 (PI: Ferraro), 195.D-0750 (PI: Ferraro),
  193.D-0232 (PI: Ferraro). This publication makes use of data
  products from the Two Micron All Sky Survey, which is a joint
  project of the University of Massachusetts and the Infrared
  Processing and Analysis Center/California Institute of Technology,
  funded by the National Aeronautics and Space Administration and the
  National Science Foundation.}}

\correspondingauthor{Cristina Pallanca}
\email{cristina.pallanca3@unibo.it}

\author[0000-0002-7104-2107]{Cristina Pallanca}
\affil{Dipartimento di Fisica e Astronomia, Universit\`a di Bologna, Via Gobetti 93/2 I-40129 Bologna, Italy}
\affil{INAF-Osservatorio di Astrofisica e Scienze dello Spazio di Bologna, Via Gobetti 93/3 I-40129 Bologna, Italy}

\author[0000-0001-9545-5291]{Silvia Leanza}
\affil{Dipartimento di Fisica e Astronomia, Universit\`a di Bologna, Via Gobetti 93/2 I-40129 Bologna, Italy}
\affil{INAF-Osservatorio di Astrofisica e Scienze dello Spazio di Bologna, Via Gobetti 93/3 I-40129 Bologna, Italy}

\author[0000-0002-2165-8528]{Francesco R. Ferraro}
\affil{Dipartimento di Fisica e Astronomia, Universit\`a di Bologna, Via Gobetti 93/2 I-40129 Bologna, Italy}
\affil{INAF-Osservatorio di Astrofisica e Scienze dello Spazio di Bologna, Via Gobetti 93/3 I-40129 Bologna, Italy}

\author[0000-0001-5613-4938]{Barbara Lanzoni}
\affil{Dipartimento di Fisica e Astronomia, Universit\`a di Bologna, Via Gobetti 93/2 I-40129 Bologna, Italy}
\affil{INAF-Osservatorio di Astrofisica e Scienze dello Spazio di Bologna, Via Gobetti 93/3 I-40129 Bologna, Italy}

\author[0000-0003-4237-4601]{Emanuele Dalessandro}
\affil{INAF-Osservatorio di Astrofisica e Scienze dello Spazio di Bologna, Via Gobetti 93/3 I-40129 Bologna, Italy}

\author[0000-0002-5038-3914]{Mario Cadelano}
\affil{Dipartimento di Fisica e Astronomia, Universit\`a di Bologna, Via Gobetti 93/2 I-40129 Bologna, Italy}
\affil{INAF-Osservatorio di Astrofisica e Scienze dello Spazio di Bologna, Via Gobetti 93/3 I-40129 Bologna, Italy}

\author[0000-0003-2742-6872]{Enrico Vesperini}
\affil{Department of Astronomy, Indiana University, Bloomington, IN, 47401, USA}

\author[0000-0002-6040-5849]{Livia Origlia}
\affil{INAF-Osservatorio di Astrofisica e Scienze dello Spazio di Bologna, Via Gobetti 93/3 I-40129 Bologna, Italy}

\author[0000-0001-9158-8580]{Alessio Mucciarelli}
\affil{Dipartimento di Fisica e Astronomia, Universit\`a di Bologna, Via Gobetti 93/2 I-40129 Bologna, Italy}
\affil{INAF-Osservatorio di Astrofisica e Scienze dello Spazio di Bologna, Via Gobetti 93/3 I-40129 Bologna, Italy}

\author[0000-0002-6092-7145]{Elena Valenti}
\affil{European Southern Observatory, Karl-Schwarzschild-Strasse 2, 85748 Garching bei Munchen, Germany}
\affil{Excellence Cluster ORIGINS, Boltzmann-Strasse 2, D-85748 Garching Bei Munchen, Germany}

\author{Andrea Miola}
\affil{Dipartimento di Fisica e Scienze della Terra,  Universit\`a di Ferrara, via Saragat 1, I-44122, Ferrara, Italy}





\begin{abstract}
In the context of a project aimed at characterizing the properties of
star clusters in the Galactic bulge, here we present the determination
of the internal kinematics and structure of the massive globular
cluster NGC 6569. The kinematics has been studied by means of an
unprecedented spectroscopic dataset acquired in the context of the
ESO-VLT Multi-Instrument Kinematic Survey (MIKiS) of Galactic globular
clusters, combining the observations from four different
spectrographs.  We measured the line-of-sight velocity of a sample of
 almost $1300$ stars distributed between $\sim 0.8\arcsec$ and
$770\arcsec$ from the cluster center.  From a sub-sample of
high-quality measures, we determined the velocity dispersion profile
of the system over its entire radial extension (from $\sim 5\arcsec$
to $\sim 200\arcsec$ from the center), finding the characteristic
behavior usually observed in globular clusters, with a constant inner
plateau and a declining trend at larger radii.  The projected density
profile of the cluster has been obtained from resolved star counts, by
combining high-resolution photometric data in the center, and the Gaia
EDR3 catalog radially extended out to $\sim 20\arcmin$ for a proper
sampling of the Galactic field background.  The two profiles are
properly reproduced by the same King model, from which we estimated
updated values of the central velocity dispersion, main structural
parameters (such as the King concentration, the core, half-mass, and tidal
radii), total mass, and relaxation times.  Our analysis also reveals a
hint of ordered rotation in an intermediate region of the cluster
($40\arcsec<r<90\arcsec$, corresponding to $ 2 r_c<r<4.5 r_c$), but
additional data are required to properly assess this possibility.
\end{abstract}



\section{Introduction}
\label{sec:intro}
The Galactic bulge is the sole spheroid where individual stars can
be observed, and for this reason is a formidable laboratory to study
the processes which drive the formation of galaxy
bulges. Unfortunately, because of observational limitations mainly
related to the large extinction and stellar density in this direction
of the sky, it still remains one of the most inaccessible regions of
the Milky Way and its structure, formation, and evolution are still
subjects of intense debate in the literature (see, for example,
\citealt{rich13,ness+13,origlia14,zoccali+16,johnson+22,marchetti+22}).  In this respect the
investigation of the globular clusters (GCs) orbiting the bulge is a
key tool to trace the bulge properties in terms of kinematics,
chemical abundances, and age \citep{bica+06, valenti+07, valenti+10,
  barbuy+18}.

For this reason several years ago we initiated a systematic
photometric and spectroscopic investigation of bulge GCs and their
stellar populations
 (see \citealt{cadelano+17a, cadelano+20, cadelano+22_6256,
 deras+23, ferraro+09, ferraro+15, leanza+22,
 origlia+97, origlia+01, origlia+02, origlia+04, origlia+05, 
 pallanca+19, pallanca+21,   pallancaDRCliller,
 saracino+15, saracino+16, saracino+19,
  valenti+05, valenti+07, valenti+10, valenti+11}).
    In this respect, the discovery that
Terzan5 and Liller1 are not genuine GCs but host multi-iron and
multi-age stellar populations has opened a new line of investigation,
providing the first detection of {\it Bulge Fossil Fragments}, the
possible remnants of primordial giant clumps that contributed to the
bulge formation, surviving the violent phase of its assembling
\citep{ferraro+09, ferraro+16,ferraro+21, lanzoni+10, origlia+11,
  origlia+13, origlia+19, massari+14, dalessandro+22}.

In the framework of a complete kinematical, chemical and photometric
characterization of bulge GCs, the ESO-VLT Multi-Instrument Kinematic
Survey (hereafter the MIKiS survey; \citealt{ferraro+18a,ferraro+18b})
is expected to provide an important contribution.  The survey has been
specifically designed to characterize the kinematical properties of a
representative sample of Galactic GCs (GGCs) in different environments
(halo and bulge) and in different dynamical evolutionary stages. The
approach proposed in MIKiS is to derive both the velocity dispersion
and the rotation profiles of the investigated systems from the
line-of-sight velocities of a statistically significant sample of
individual stars distributed over their entire radial extension.  The
survey was designed to exploit the remarkable performances of the
spectroscopic capabilities currently available at the ESO Very Large
Telescope (VLT). In particular, it uses the adaptive-optics (AO)
assisted integral-field spectrograph SINFONI (operating in the
infrared) and MUSE (in the optical band), the multi-object integral
field spectrograph KMOS, and the multi-object, fiber-fed and
wide-field spectrograph FLAMES/GIRAFFE. The dataset has been collected
in the framework of three Large Programs (namely 193.D-0232,
195.D-0750, 106.21N5, PI: Ferraro; the last one is still ongoing),
complemented with a series of specific proposals. Some recent results
can be found in \citet{ferraro+18a, lanzoni+18a, lanzoni+18b,
  leanza+22b}.  When possible this approach is also complemented with
accurate measures of individual proper motions (PMs; see
\citealt{massari+13, cadelano+17b, raso+20, libralato+18,
  libralato+22}).

In this paper, we present the velocity dispersion profile of NGC 6569,
a massive GC (with absolute total magnitude $M_V=-8.3$) located in the
Sagittarius region ($l= 0.48\arcdeg, b=-6.68\arcdeg$; Harris 1996,
2010 edition) of the Galactic bulge, at a distance of $\sim 3$ kpc
from the Galactic center (Harris 1996). The cluster is projected
toward the dark nebula Barnard 305 \citep{Barnard+27} and it is
therefore highly reddened, with an average color excess $E(B-V)=0.53$
\citep{Ortolani+01}. It is an intermediate-high metal rich cluster,
with a quoted metallicity ranging from [Fe/H]$= -0.79\pm 0.02$
\citep{valenti+11} to [Fe/H]$= -0.87$ \citep{johnson+18}, and with an
$\alpha-$element enhancement of $[\alpha/$Fe$]=+0.4$ \citep{valenti+11} . 
This system has
been subject to a detailed photometric analysis by our group (see
\citealt{saracino+19}) by using a combination of optical Hubble Space
Telescope/WFC3 data, and multi-conjugate adaptive optics assisted
near-infrared (NIR) GEMINI observations. This allowed an accurate
measure of PMs, which provided a robust selection of cluster member
stars. A differential reddening map has been also derived. The
PM-selected and differential reddening CMD then has allowed an
accurate measure of the cluster distance and age: the distance modulus
is $(m-M)_0=15.03\pm0.08$, corresponding to $10.1\pm0.2$ Kpc from the
Sun, while the age amounts to about 12.8 Gyr, with an uncertainty of
$0.8-1.0$ Gyr \citep{saracino+19}.
 
The paper is organized as follows. In Section 2 we present the
observations and describe the procedures adopted for the data
reduction. In Section 3 we discuss the selection of the samples, the
methods to determine the radial velocities (RVs), and the strategy
adopted to homogenize the different datasets available. The results
are presented in Section 4, while Section 5 is devoted to the
discussion and conclusions.

\section{Observations and data reduction}
\label{sec:obs}
As anticipated in the Introduction, to build the velocity dispersion
profile of NGC 6569 we used a multi-instrument approach combining
the RV measurements obtained from four different spectroscopic
datasets.
\medskip\\ \textit{MUSE/NFM - } The innermost cluster regions were
sampled mainly by using the AO-assisted integral-field spectrograph
MUSE in the Narrow Field Mode \citep[NFM,][]{bacon+10}. This is the
MUSE configuration that provides the highest spatial resolution.  MUSE
at ESO-VLT consists of a modular structure composed of 24 identical
Integral Field Units (IFUs) and, in the NFM configuration, it is
equipped with the Adaptive Optics Facility (AOF) of the VLT and the
GALACSI-AO module \citep{Arsenault+08, Strobele+12}.  At the highest
spatial sampling ($0.025\arcsec$/pixel), MUSE/NFM observations cover a
field of view of $7.5\arcsec\times 7.5\arcsec$, which is smaller than
that provided by the wide field mode (WFM) configuration
($1\arcmin\times 1\arcmin$, with a sampling of $0.2\arcsec$/pixel).
MUSE provides a wavelength coverage from $4800$ \AA\ to $9300$ \AA,
with a resolving power R $\sim3000$ at $\lambda\sim8700$ \AA.  Our
dataset has been collected as part of the ESO Large Program ID:
106.21N5.003 (PI: Ferraro; see Table \ref{tab:data}) and consists of a
mosaic of seven MUSE/NFM pointings sampling the innermost $\sim
15\arcsec$ from the center.  Three 750 s long exposures were acquired
for each pointing, with an average DIMM seeing always better than
$\sim0.7\arcsec$.  A small dithering pattern and a rotation offset of
$90 \arcdeg$ were secured between consecutive exposures of the same
pointing, in order to remove possible systematic effects between the
individual IFUs.  The data reduction was performed with the dedicated
MUSE ESO pipeline \citep{Weilbacher+20}. It performs the basic
reduction (bias subtraction, flat fielding, and wavelength
calibration) for each individual IFU, then, applies the sky
subtraction and transforms the pre-processed data into physical
quantities by performing the flux and astrometric calibration for each
IFU, and applying the heliocentric velocity correction to all the
data.  In the next step, the data from all 24 IFUs are combined into a
single datacube, and, finally, the pipeline combines the datacubes of
the multiple exposures of each pointing into a final single datacube,
taking into account the dithering offsets and rotations among
different exposures.
\medskip\\ \textit{SINFONI -} To sample the innermost regions, we also
used an additional spectroscopic dataset acquired with the
near-infrared ($1.1 - 2.45 \mu$m) AO-assisted integral-field
spectrograph SINFONI \citep{Eisenhauer+03} at the ESO-VLT.  The
observations (ESO Large program ID: 195.D-0750(A), PI: Ferraro) were
performed with the $K-$band grating, which samples the wavelength
range $1.95 - 2.45 \ \mu$m and provides a spectral resolution R $\sim$
4000, with a spatial scale of $0.25\arcsec$/spaxel corresponding to a
field of view of $8\arcsec\times 8\arcsec$.  The dataset is listed in
Table \ref{tab:data} and includes eight pointings covering a region of
$\sim15\arcsec$ from the cluster center.  For each pointing, multiple
exposures (usually six) were acquired on the target and on a sky
region following a typical target-sky sky-target sequence, in order to
allow an adequate subtraction of the background.  The observations
have been executed adopting an exposure time of 30 s and under an
average DIMM seeing of $\sim 0.8\arcsec$.  The data reduction was
performed by using the standard ESO pipeline (esorex 3.13.6) following
the workflow 3.3.2 under the EsoReflex environment
\citep{Freudling+13}.  In a first step, the pipeline applies the
corrections for darks, flats, and geometrical distortions to all the
target and sky exposures. Then, it subtracts the sky background and
performs the wavelength calibration.  Finally, the processed target
frames are combined in a datacube for each exposure.
\medskip\\ \textit{KMOS - } The cluster region at intermediate
distances from the center has been sampled by using the integral-field
spectrograph KMOS \citep{Sharples+13} at the ESO-VLT.  KMOS employs 24
IFUs, each one with a field of view of $2.8\arcsec\times 2.8\arcsec$
and a spatial sampling of $0.2\arcsec$/pixel. The IFUs can be
allocated within a 7.2$\arcmin$ diameter field of view.  The data have
been acquired under the ESO Large Program ID: 193.D-0232 (PI: Ferraro)
adopting the YJ grating, which samples the 1.025-1.344 $\mu$m spectral
range at a resolution R $\sim$ 3400, and with the spectral sampling of
$\sim 1.75$ \AA/pixel.  Nine pointings (see Table \ref{tab:data}) were
acquired within $\sim 3\arcmin$ from center.
Four pointings consist of a sequence of three repeated sub-exposures
each one 60s long, the other five pointings are a set of three longer
(100s-long) repeated sub-exposures that were acquired to sample
fainter stars.  The spectroscopic targets have been selected from the
near-infrared catalog
described in \citet{valenti+05, valenti+07}
\footnote{The catalog is available at the web site
\url{http://www.bo.astro.it/~GC/ir_archive/}}, complemented, in the
outermost regions of the clusters, with data from the 2MASS catalog
\citep{2mass}.
Typically, the KMOS observations were planned to have one/two red
giant branch (RGB) stars in the field of view of each IFU (see also
\citealt{lapenna+15}).  The data reduction has been performed by
adopting the dedicated ESO
pipeline,\footnote{\label{note1}http://www.eso.org/sci/software/pipelines/}
executing background subtraction, flat field correction, and
wavelength calibration.
\medskip\\ \textit{FLAMES - } To investigate the cluster kinematics in
the outermost cluster regions we used the fiber-fed multi-object
spectrograph FLAMES \citep{Pasquini+02} in the GIRAFFE/MEDUSA mode,
which consists of 132 fibers with an aperture of 1.2$\arcsec$ each.
The observations have been executed adopting the HR21 and HR13 grating
setups (ESO Large Program ID: 193.D-0232(F), PI: Ferraro, and ID:
093.D-0286(A), PI: Villanova, respectively; see Table \ref{tab:data}).
The HR13 dataset samples the spectral range $6120 - 6405$ \AA \ with a
resolving power $R\sim 26400$, and consists of two repeated exposures
(each long 2775 s) of the same targets.  The HR21 grating provides a
resolving power $R\sim 18000$ sampling the wavelength range between
8484 and 9001 \AA.  The targets have been selected to sample the full
extension in luminosity of the RGB. Thus, 6 pointings were planned to
optimize the observations of targets with different luminosities: the
brightest portion of the RGB has been sampled through three pointings
(each with a 1800 s long exposure), while two pointings with an
exposure time of 2700 s were devoted to observe
intermediate-luminosity RGB stars, and finally an additional pointing
of 5400 s was dedicated to sample the fainter portion of the RGB.  As
for the KMOS dataset, the targets have been selected from the
SOFI/2MASS photometric catalogs by sampling the full  extension along the
cluster RGB, and the data have been reduced with the dedicated ESO
pipelines.$^2$

\begin{deluxetable*}{clcCC}
\tablecaption{Spectroscopic datasets for NGC 6569.}
\tablewidth{0pt}
\setlength{\tabcolsep}{10pt} 
\renewcommand{\arraystretch}{1} 
\tablehead{
\colhead{ Name }  &  \colhead{}  &
\colhead{Date} & \colhead{N$_{\rm exp}$} & \colhead{$t_{\rm exp}$ [s]}  }
\startdata
\hline
\multicolumn{5}{c}{MUSE/NFM}\\
\hline
C & & 2021-08-16 & 3 & 750\\
E & & 2021-08-16 & 3 & 750\\
N & & 2021-08-16 & 3 & 750\\
NE & & 2021-08-21 & 3 & 750\\
W  & & 2021-08-21 & 3 & 750\\
SW & & 2021-08-21 & 3 & 750\\
S  & & 2022-09-27 & 3 & 750\\
\hline
\multicolumn{5}{c}{SINFONI}\\
\hline
LR\TextUnderscore{}SW  & & 2015-08-02 & 6 & 30\\
LR\TextUnderscore{}W & & 2015-08-03 & 6 & 30\\
LR\TextUnderscore{}N & & 2015-07-18 & 6 & 30\\
LR\TextUnderscore{}NE & & 2016-07-22/2016-08-04 & 6 & 30\\
LR\TextUnderscore{}S & & 2016-06-23 & 6 & 30\\
LR\TextUnderscore{}NN & & 2016-07-15 & 6 & 30\\
LR\TextUnderscore{}C & & 2015-08-23 & 6 & 30\\
LR\TextUnderscore{}E & & 2016-07-21 & 12 & 30\\
\hline
\multicolumn{5}{c}{KMOS}\\
\hline
kmos\TextUnderscore{}1 & & 2015-04-30 & 3 & 60 \\
kmos\TextUnderscore{}2 & & 2015-05-02 & 3 & 60 \\
kmos\TextUnderscore{}3 & & 2015-05-02 & 3 & 60 \\
kmos\TextUnderscore{}4 & & 2015-05-02 & 3 & 60 \\
kmos\TextUnderscore{}faint\TextUnderscore{}1 & & 2015-05-02 & 3 & 100 \\
kmos\TextUnderscore{}faint\TextUnderscore{}2 & & 2015-05-02 & 3 & 100 \\
kmos\TextUnderscore{}faint\TextUnderscore{}3 & & 2015-05-02 & 3 & 100 \\
kmos\TextUnderscore{}faint\TextUnderscore{}4 & & 2015-05-03 & 3 & 100 \\
kmos\TextUnderscore{}faint\TextUnderscore{}5 & & 2015-05-02 & 3 & 100 \\
\hline
\multicolumn{5}{c}{FLAMES}\\
\hline
flames\TextUnderscore{}HR13 & & 2014-06-19/2014-08-01 & 2 & 2775 \\
flames\TextUnderscore{}HR21\TextUnderscore{}faint\TextUnderscore{}1 & & 2015-06-22 & 1 & 1800\\
flames\TextUnderscore{}HR21\TextUnderscore{}faint\TextUnderscore{}2 & & 2015-07-26 & 1 & 1800 \\
flames\TextUnderscore{}HR21\TextUnderscore{}faint\TextUnderscore{}3 & & 2015-07-26 & 1 & 1800 \\
flames\TextUnderscore{}HR21\TextUnderscore{}veryfaint\TextUnderscore{}1 & & 2015-06-27 & 1 & 2700 \\
flames\TextUnderscore{}HR21\TextUnderscore{}veryfaint\TextUnderscore{}2 & & 2015-06-27 & 1 & 2700 \\
flames\TextUnderscore{}HR21\TextUnderscore{}VVfaint\TextUnderscore{}2 & & 2015-06-28/2015-07-26 & 2 & 2700 \\
\hline \enddata
\tablecomments{For each of the datasets analyzed in this work
  (MUSE/NFM, SINFONI, KMOS and FLAMES), and for each individual
  pointing, the table lists the name, execution date, number of
  exposures (N$_{\rm exp}$) and exposure time of each frame ($t_{\rm
    exp}$, in seconds).}
\label{tab:data}
\end{deluxetable*}

\newpage
\section{Center of Gravity}
\label{sec:centro}
For a proper analysis of the density and radial velocity distributions
of stars in NGC 6569, the first step is the determination of the
center of gravity of the cluster.  To this aim, we used the
photometric catalog described in \citet{saracino+19}.  This is based
on F555W and F814W images obtained from high-resolution HST/WFC3
observations, and a set of $J$- and $K_s$-band images acquired with
the Gemini multi-conjugate adaptive optic system (GeMS). 
 The field of view of the GeMS observations is almost entirely included within that of the HST/WFC3 images, 
which extend out to $\sim 150\arcsec$  from the center \citep[see Figure 2 of ][]{saracino+19}. 
As in previous papers, we have determined the position of the
gravitational center ($C_{\mathrm{grav}}$) from the position of
resolved  stars (in the HST data-set), rather than the surface brightness peak. This is done
to avoid possible biases induced by the presence of a few bright
stars, which could significantly offset the location of the surface
brightness peak, with respect to the real $C_{\mathrm{grav}}$.

To identify the position of $C_{\mathrm{grav}}$, we adopted the
iterative procedure described in \citet{montegriffo+95} and used in
many other papers \citep[see, e.g.][]{lanzoni+10, lanzoni+19,miocchi+13, cadelano+20,pallanca+21}.  
This method computes the position of
$C_{\mathrm{grav}}$ by averaging the projected coordinates ($x$, $y$)
on the plane of the sky of a sample of resolved stars selected in an
appropriate range of magnitude and within a given radial distance
($r$) from the center, starting from a first-guess value of the
latter.  Among the targets of the photometric catalog described in
\cite{saracino+19}, we considered only the stars brighter than
$\rm m_{F814W}=20.0,\ 20.3,\ 20.6$,
which are reasonable selections to obtain statistically large samples 
while avoiding incompleteness effects.  For each magnitude cut, we
considered the stars included within circles of different radii
($r=30\arcsec$, $r=34\arcsec$, $r=38\arcsec$) from the adopted
center. These values are larger than the cluster core radius quoted in
the literature ($r_c = 21\arcsec$; \citealp{harris+96}) to ensure that
the procedure is applied in a region where the density profile
starts to decrease \citep[see][]{miocchi+13}.  As first-guess center
we adopted the value quoted in \citet{harris+96}.  Then, from each
sub-sample of selected stars we computed a new guess value of the
cluster center by averaging the stellar coordinates projected on the
plane of the sky. The procedure is repeated iteratively by using each
time the center value computed in the previous iteration, until
convergence.  The convergence is reached when ten consecutive
iterations provide values of the center that differ by less than
$0.01\arcsec$ from each other.  We determined the final position of
$C_{\mathrm{grav}}$ of NGC 6569 as the average of the values obtained
from each sub-sample, finding $\alpha = 18^{\mathrm{h}}
13^{\mathrm{m}} 38.70^{\mathrm{s}}$, $\delta = -31\arcdeg 49\arcmin
37.13\arcsec$, with an uncertainty of $\sim 0.3\arcsec$.  This is
located at $\sim0.1\arcsec$ West and $\sim0.3\arcsec$ South from the
previous estimate reported in \citet{harris+96}.  In the next
analysis, we have always adopted the position of the cluster center
obtained in this work.

\section{Radial Velocity measurements}
\label{sec:RV}
To properly derive the RV of individual stars from the spectra
acquired with the four different spectrographs used in this work, we
have performed a specific analysis of each dataset following a
procedure similar to the one described in \citet{leanza+22b}.  Below
we summarize only the main steps.

\begin{figure}[ht!]
\centering
\includegraphics[width=\textwidth]{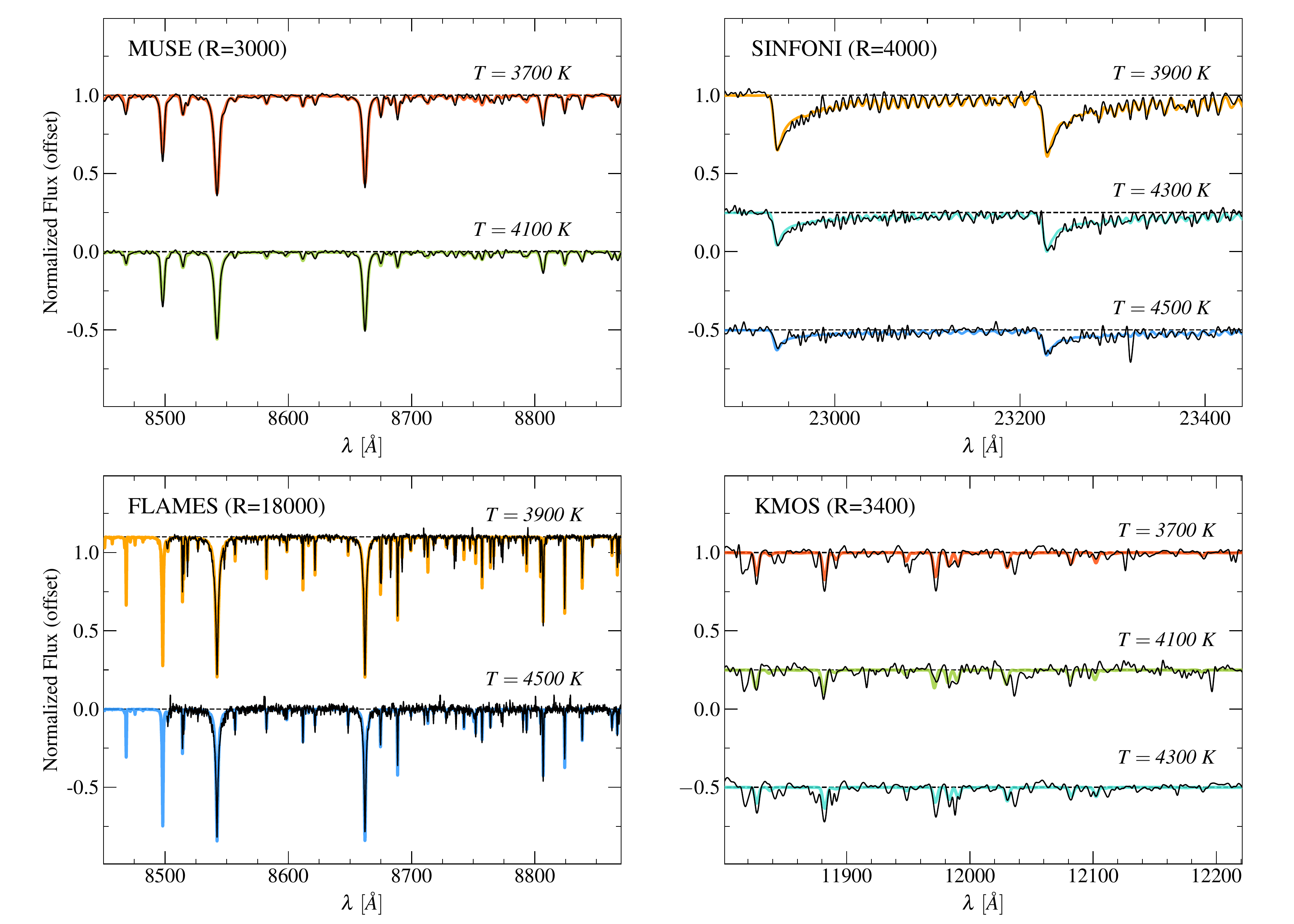}
\centering
\caption{Examples of normalized observed spectra (in black) of stars
  with different atmospheric parameters, acquired with MUSE/NFM
  (top-left), SINFONI (top-right), FLAMES in the HR21 setup
  (bottom-left) and KMOS (bottom-right).
  For each observed spectrum, the best-fit synthetic spectrum obtained
  from the procedures described in Section \ref{sec:RV} and shifted by
  the derived RV value is overplotted in a color code, according to the effective temperature (labelled for each spectrum) associated to the best-fit template. The
  difference in the spectral resolution (labelled in each panel) of the various instruments is
  evident, as it is also the temperature dependence of the CO
  band-head deepness in the SINFONI spectra. }
\label{fig:spectra}
\end{figure}

\begin{figure}[ht!]
\centering
\includegraphics[width=\textwidth]{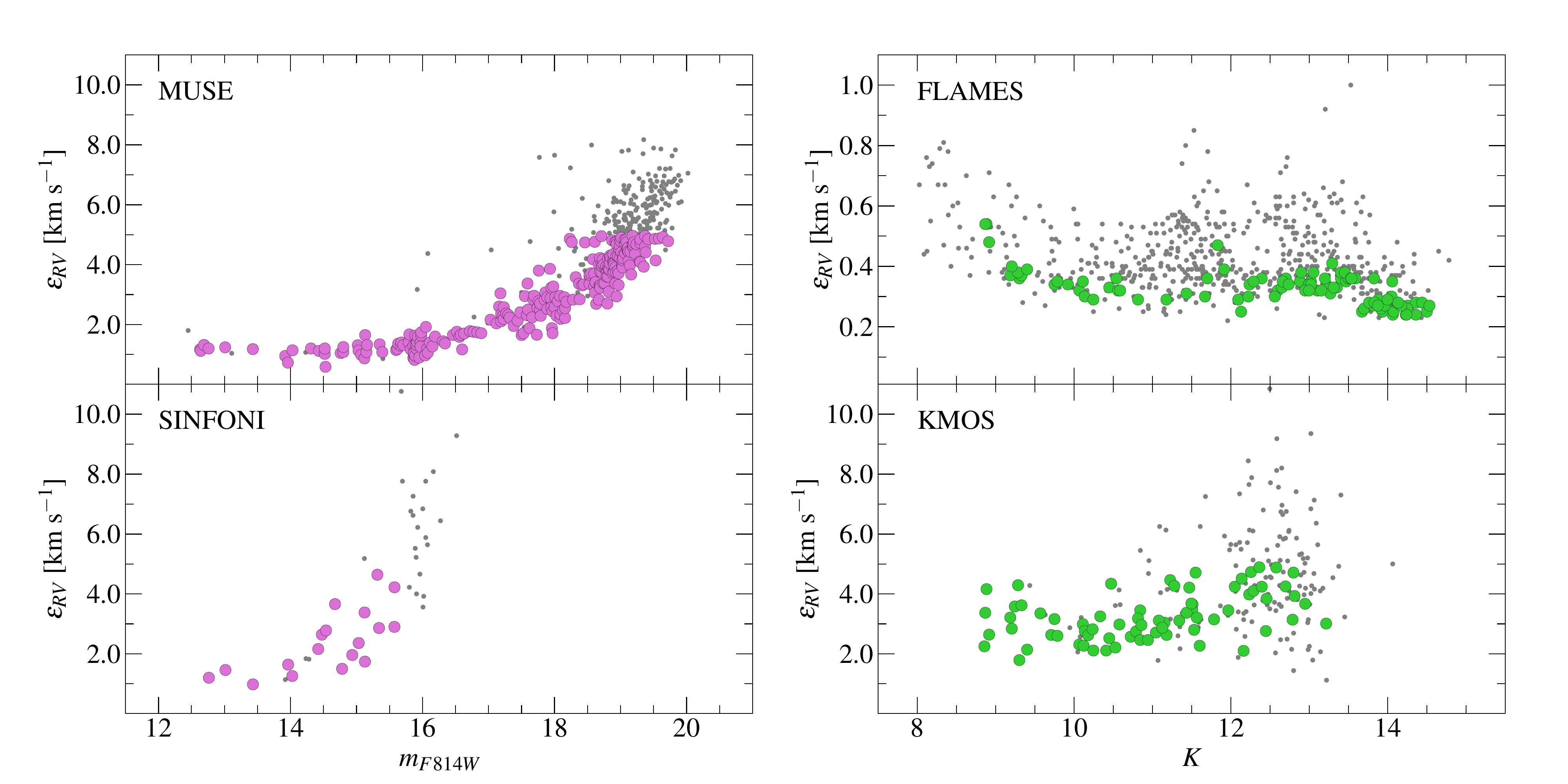}
\centering
\caption{RV uncertainty ($\epsilon_{\rm RV}$) as a function of the
  star magnitude for the stars observed in the MUSE/NFM, SINFONI,
  FLAMES, and KMOS datasets (top-left, bottom-left, top-right, and
  bottom-right panels, respectively; see labels).  In all panels, the
  colored circles correspond to the targets surviving the membership
  and the quality selections (see Section \ref{sec:kin}), while the
  gray dots are the stars rejected from the analysis.   }
\label{fig:err_mag}
\end{figure}

\begin{figure}[ht!]
\centering
\includegraphics[width=\textwidth]{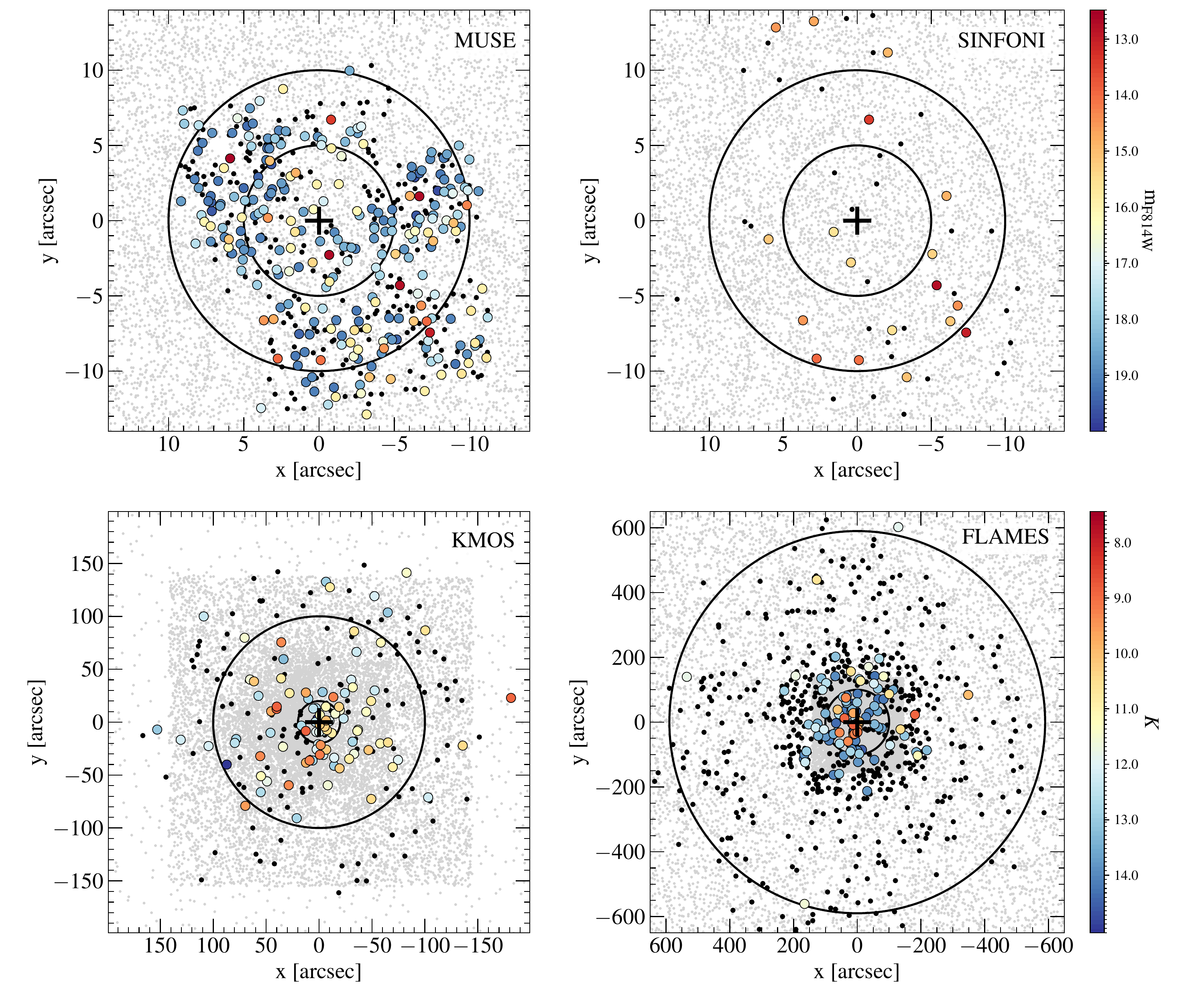}
\centering
\caption{Maps in the plane of the sky, with respect to the adopted
  cluster center (black cross, determined in Section
  \ref{sec:centro}), of the stars with measured RV in each dataset.
  In all panels, the colored large circles mark the targets that
  survived the membership and quality selections and have been used
  for the kinematic analysis (see Section \ref{sec:kin}), while the
  black dots are the rejected stars.  The color scales indicate the
  $\rm m_{F814W}$ and $K$-band magnitudes for the inner sample (MUSE and
  SINFONI, top panels) and for the external datasets (KMOS and FLAMES,
  bottom panels), respectively.  \textit{Top-left}: focus on the
  central region sampled by MUSE.  The gray dots are the stars of the
  HST/GeMS photometric catalog \citep{saracino+19}.  The two circles
  mark distances of 5$\arcsec$ and $10\arcsec$ from the center.
  \textit{Top-right}: as in the top-left panel but relating to the
  SINFONI dataset.  \textit{Bottom-left}: map relative to the position
  of the KMOS targets.  The gray dots in the background mark the stars
  in the SOFI/2MASS catalog.  The two circles are centered in the
  cluster center and have radii of 19.9$\arcsec$ (equal to the core
  radius of the cluster; see Section \ref{sec:discussion}) and
  $100\arcsec$.  \textit{Bottom-right}: external portion of the
  cluster sampled by the FLAMES targets. The gray dots are as in the
  bottom-right panel, while the two circles mark distances of
  100$\arcsec$ and 589.7$\arcsec$ (corresponding to the truncation
  radius of the cluster; see Section \ref{sec:discussion}) from the
  center. }
\label{fig:mappa}
\end{figure}

{\it MUSE -} The MUSE spectra have been extracted by using the
software PampelMuse \citep{kamann+13}, which allows us to obtain
deblended source spectra of individual stars even in crowded regions
of stellar systems, by performing a wavelength-dependent point spread
function (PSF) fitting. PampelMuse uses as input a reference catalog
with the magnitudes and the coordinates of all the stars present in
the field of view of the datacube. For this purpose we adopted the
photometric catalog obtained by \cite{saracino+19}.  As PSF model we
selected the MAOPPY function \citep{fetick+19}, which was developed to
properly reproduce both the core and the halo of the AO-corrected PSF
in MUSE/NFM observations \citep[for more details see][]{gottgens+21}.
Briefly, for each slice of the MUSE datacube, PampelMuse fits the PSF
and a coordinate transformation from the reference catalog to the
data, and uses these quantities to extract the spectra of all the
stars in the datacube optimizing the deblending of the sources. The RV
measures have been derived from the Doppler shifts of the Calcium
Triplet lines in the wavelength range $8450-8750$ \AA \ by following
the same procedure described in the previous paper \citet{leanza+22b}.
To this aim, a library of synthetic spectra computed with the SYNTHE
code 
has been used.  The template
spectra have been produced in the wavelength range covered by MUSE,
adopting the cluster metallicity with an $\alpha-$enhanced chemical
mixture ([Fe/H]$=-0.79$ dex and [$\alpha$/Fe]$=0.4$ dex, respectively,
\citealt{valenti+11}), appropriate atmospheric parameters (effective
temperature and gravity), according to the evolutionary phase of the
targets, and applying a convolution with a Gaussian profile to obtain
the MUSE spectral resolution.  Briefly, the procedure, firstly,
normalizes the spectra corrected for heliocentric velocity to the
continuum, which is estimated by a spline fitting of the spectrum in
an appropriate wavelength range.  Then, it computes the residuals
between the normalized observed spectra and each template of the
library, shifted in RV by steps of 0.1 km s$^{-1}$ in an appropriate
velocity range.  By determining the smallest standard deviation in the
distribution of the residuals, the procedure then provides, as a
result, the best-fit synthetic spectrum (hence, the best estimates of
temperature and gravity), and the RV of the target.  An estimate of
the signal to noise ratio (S/N) of the spectrum is also computed as
the ratio between the average of the counts and their standard
deviation in the wavelength range $8000 - 9000$ \AA.
The top-left panel of figure \ref{fig:spectra} shows an example of the
output of this procedure.  For stars with different atmospheric
parameters (the estimated effective temperature is labelled), and in
the wavelength range used to estimate the RV, each panel shows the
observed spectrum (in black) and the best-fit synthetic spectrum
shifted by the estimated RV (in a color code).
The RV uncertainties have been determined by running 9000 Monte Carlo
simulations of spectra with S/N between 10 and 90, and applying to
these simulated spectra the same procedure used for the observed ones
\citep[for more details see][]{leanza+22b}.  The typical RV errors are
lower than 2 km s$^{-1}$ for the brightest stars, and increase as
function of the magnitude up to $\sim 8$ km s$^{-1}$, as shown in
Figure \ref{fig:err_mag} (top left panel). In the case of overlapping
MUSE pointings, the stars in common have been used to search for
possible systematic offsets in the RV measures,
always finding a good agreement within the errors.
In the case of stars with multiple exposures, we performed a weighted
mean of all the RV measures, by using the individual errors as
weights. In total we obtained RV measures for a sample of 475
targets. The position of these targets in the plane of the sky is
plotted in Figure \ref{fig:mappa}, while the color scale shows the
magnitude range covered by the sample.

{\it SINFONI -} The method adopted for the extraction and the analysis
of the SINFONI spectra has been shortly presented in
\citet{leanza+22b} and will be extensively described in a forthcoming
paper (C. Pallanca et al., 2023, in preparation).  SINFONI
spectra have been extracted from all the spaxels showing photon counts
above a fixed threshold that was assumed at $10 \sigma$ from the
background level. RVs were then computed by following a procedure
analogous to that applied for the MUSE spectra.  In the case of
SINFONI, we used the $^{12}C^{16}O$ band-heads as reference lines to
determine the Doppler shift.  The set of synthetic spectra has been
computed with the SYNTHE code \citep{Sbordone+04, kurucz+05} adopting
temperature and surface gravity typical of RGB stars, metallicity and
$\alpha-$element abundances of the cluster \citep{valenti+11}.  The
synthetic spectra have been computed to cover the same NIR wavelength
range covered by the SINFONI observations with the same spectral
resolution (by means a convolution with a Gaussian profile).
Moreover, since the deepness of the CO band-heads severely depends on
both the chemical abundance and the temperature, as shown by the
example in the top-right panel of Figure \ref{fig:spectra}, and stars
above the RGB bump could be depleted in carbon, we computed 7
additional synthetic spectra with appropriate carbon-depletion
[C/Fe]$=-0.27$ dex \citep[][]{valenti+11} and different values of the
atmospheric parameters to properly reproduce the brightest portion of
the RGB, above the RGB-bump.  As mentioned above, to obtain the RVs we
applied the same procedure used for the MUSE data and also the RV
uncertainties have been estimated from similar Monte Carlo
simulations.  The errors obtained are of the order of 2 km s$^{-1}$
and also in this case they display a trend with the magnitude (see the
bottom-left panel of Figure \ref{fig:err_mag}).  No significant
offsets have been detected by comparing the RV measures of stars in
common between different overlapping pointings. At the end of this
procedure we merged the measures obtained from all the pointings by
adopting as final value of RV for each star the weighted average of
all the multiple measures using the estimated errors as weights.
Since the SINFONI dataset covers the overcrowded innermost regions of
the cluster and no deblending procedures similar to those implemented
in PampelMuse are available in this case, it is important to evaluate
the effect on the RV measures of the possible contamination of the
spectra from the light coming from bright neighboring stars. For this
reason, for each target we estimate the contamination parameter
\citep[$C$, see][]{leanza+22}. This is defined as the ratio between
the fraction of ``contaminating light'' with respect to the
contribution of the target itself in the central spaxel. The
``contaminating light'' is estimated as the expected photon counts
from the neighbouring sources providing the largest contribution to
the central spaxel and it is estimated on the basis of the PSF model
that best reproduces the SINFONI data and the list of stellar sources
from the HST/GeMS catalog.  In order to select only the safest targets
with negligible contamination from the light of neighboring sources,
we selected only the SINFONI targets with $C<0.03$. 
  The final
sample consists of 51 targets, which are all bright stars.
 
{\it FLAMES -} For the FLAMES spectra, we determined the RVs from the
Doppler shift of atomic lines in the wavelength ranges $6200-6350$
\AA \ and $8520-8870$ \AA \ for the HR13 and HR21 datasets,
respectively.  Also in this case, the library of template spectra has
been computed with the SYNTHE code \citep{Sbordone+04, kurucz+05} in
the appropriate wavelength range adopting the cluster metallicity and
different atmospheric parameters sampling the entire RGB extension.
The RV uncertainties have been derived as in the previous cases, by
means of Monte Carlo simulations, obtaining typical errors of the
order of 0.5 km s$^{-1}$ thanks to the higher spectral resolution.
The top-right panel of Figure \ref{fig:err_mag} shows that the RV
errors have a roughly constant trend as a function of the magnitude of
the stars in place of the typical increasing trend seen for the other
datasets. This is because faint stars have been observed with longer
exposure times, thus keeping the S/N ratio (and the uncertainties)
almost constant at all magnitudes.
We obtained a final FLAMES sample of 680 RV measures.  The position of
the target stars in the plane of the sky is shown in the bottom-right
panel of Figure \ref{fig:mappa}, in which the color scale represents
the magnitude.

{\it KMOS -} For the KMOS dataset, the RVs have been measured by
adopting the procedure described in \citet{lapenna+15, ferraro+18a,
  ferraro+18b}.  The spectra have been extracted from the central and
most exposed spaxel of each target star identified in the field of
view of each IFU.  Then, by using the FXCOR task under the software
IRAF, the spectra corrected for heliocentric velocity have been
cross-correlated
with appropriate synthetic spectra, according to the method described
in \citet{tonry_davis_79}.  As for the other datasets, the synthetic
spectra have been obtained with the SYNTHE code \citep{Sbordone+04,
  kurucz+05} using the wavelength range and the spectral resolution
adequate for KMOS, and the uncertainties have been computed using
similar Monte Carlo simulations.  The derived errors are of about 3 km
s$^{-1}$ and they show the trend with magnitude plotted in the
bottom-right panel of Figure \ref{fig:err_mag}.
 As described above, consistent methods have been used to measure the RVs of the MUSE, SINFONI and FLAMES targets, 
while a cross-correlation technique has been adopted in the case of KMOS spectra. 
Hence, to ensure that this introduced no systematics, we re-determined the RVs of a subsample of control stars observed with MUSE, SINFONI and FLAMES 
by using the cross-correlation implemented in IRAF. 
In all cases, we obtained RV values in excellent agreement with the previous determinations.
The final KMOS catalog consists of 220
RV measures.  The bottom-left panel of Figure \ref{fig:mappa} shows
the targets position on the plane of the sky with respect the cluster
center and the sampled magnitude range.

\subsection{Final catalog}
\label{sec:catalog}
To produce a homogeneous final catalog, we checked for possible
systematic offsets in RV among the different spectroscopic
datasets. 
Adopting the FLAMES RVs as reference, because of the highest
spectra resolution of this instrument, we compared the RV values of
the stars in common between each pair of datasets using only the most
reliable measures. We detected and applied the following offsets to
realign all the measures on the FLAMES values:  shift of $-2.4$ km
s$^{-1}$ has been applied to all the KMOS RVs, $1.9$ km s$^{-1}$ to
the MUSE values, and $-0.5$ km s$^{-1}$ to the SINFONI measures.
Moreover, to check whether the RV uncertainties are properly
estimated, we used the velocity measures ($v_1$ and $v_2$) of the
targets observed in multiple datasets and their associated errors
($\epsilon{_1}$ and $\epsilon{_2}$) to derive the quantity
\begin{equation}
 \delta v_{1,2} = \frac{v_1 - v_2} {\sqrt{\epsilon{_1}{^2} + \epsilon{_2}{^2}}},
\end{equation}
which should   return a normal distribution with a standard deviation of 1
in the case of correct uncertainties \citep[see also][]{kamann+16}.
To have a large enough sample of repeated measures, we used the stars
in common between FLAMES and KMOS, and between MUSE and SINFONI,
obtaining in both cases distributions consistent with a normal
function with a standard deviations of $\sim 1$, thus guaranteeing
that the errors are correctly estimated.

\begin{figure}[ht!]
\centering
\includegraphics[width=0.9\textwidth]{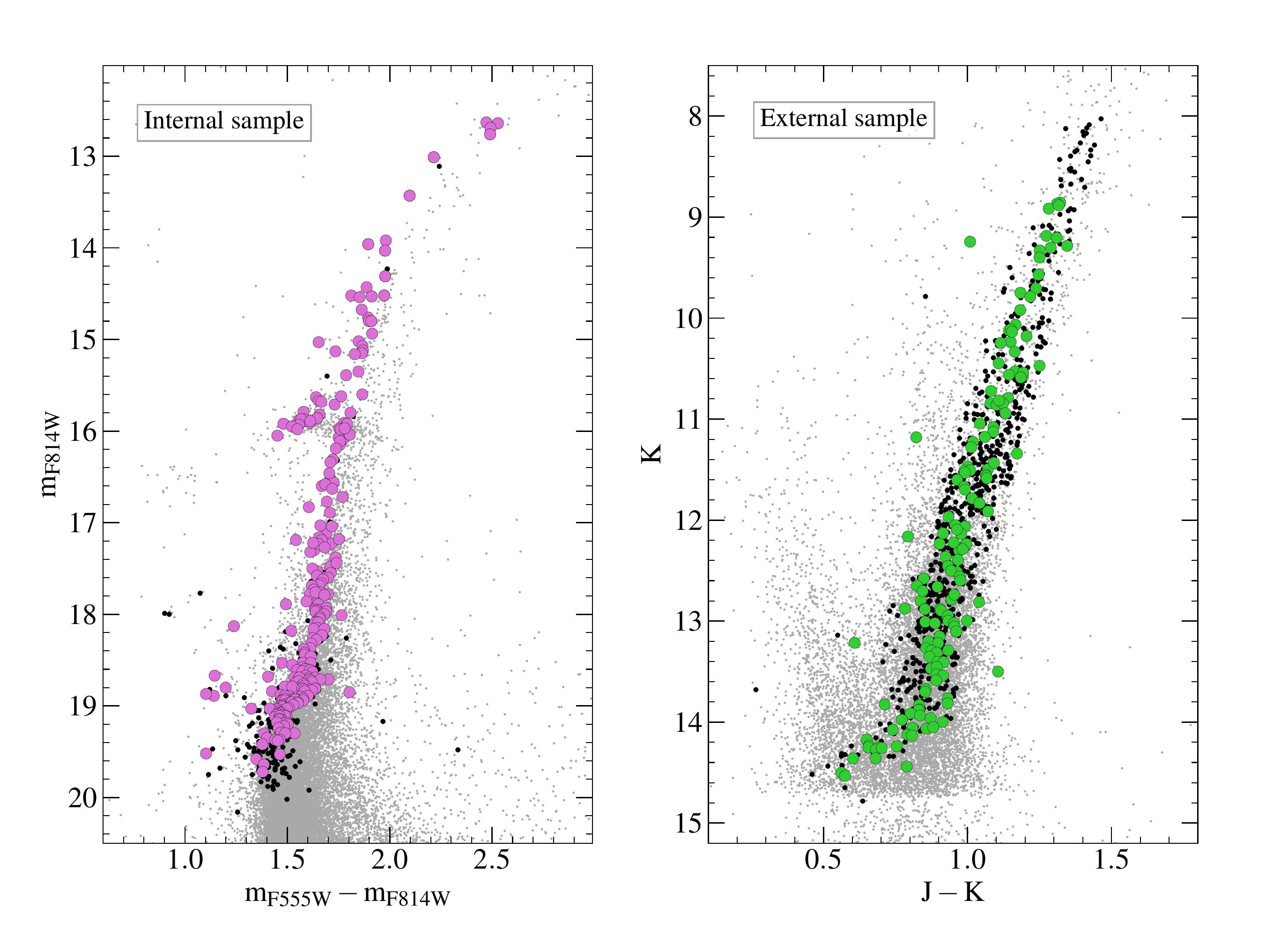}
\centering
\caption{CMDs of NGC 6569 showing the targets of the final kinematic
  catalog.  \textit{Left panel}: HST optical CMD \citep[gray dots,
    from][]{saracino+19} with the targets of the internal sample (MUSE
  and SINFONI) highlighted.  The large magenta  circles mark the targets
  after all the membership and quality selections described in
  Sections \ref{sec:kin}, while the black dots are the rejected
  targets.  \textit{Right panel}: the gray dots show the ($K, J-K$)
  CMD obtained from the SOFI/2MASS catalog. The large green circles and
  black dots are, respectively, the targets of the external sample
  (KMOS and FLAMES) selected for the kinematic analysis and rejected
  on the basis of the adopted membership and quality selections. }
\label{fig:cmd}
\end{figure}

Then, to create the final catalog, we combined the four datasets
previously homogenized by averaging the RV values of the targets with
multiple measures
using the estimated errors as weights.  We obtained a final sample of
1292 RVs of individual stars distributed from $0.8\arcsec$ to
$723.7\arcsec$ from the cluster center (corresponding to $\sim 1.2$
times the truncation radius $r_t=589.7\arcsec$; see Section
\ref{sec:discussion}).  Figure \ref{fig:cmd} shows the position of the
targets in the optical and NIR CMDs for the internal (MUSE and
SINFONI) and external (FLAMES and KMOS) samples, respectively.

\section{Internal kinematics}
\label{sec:kin}
\subsection{Cluster membership}
\label{sec:membership}
\begin{figure}[ht!]
\centering
\includegraphics[width=0.8\textwidth]{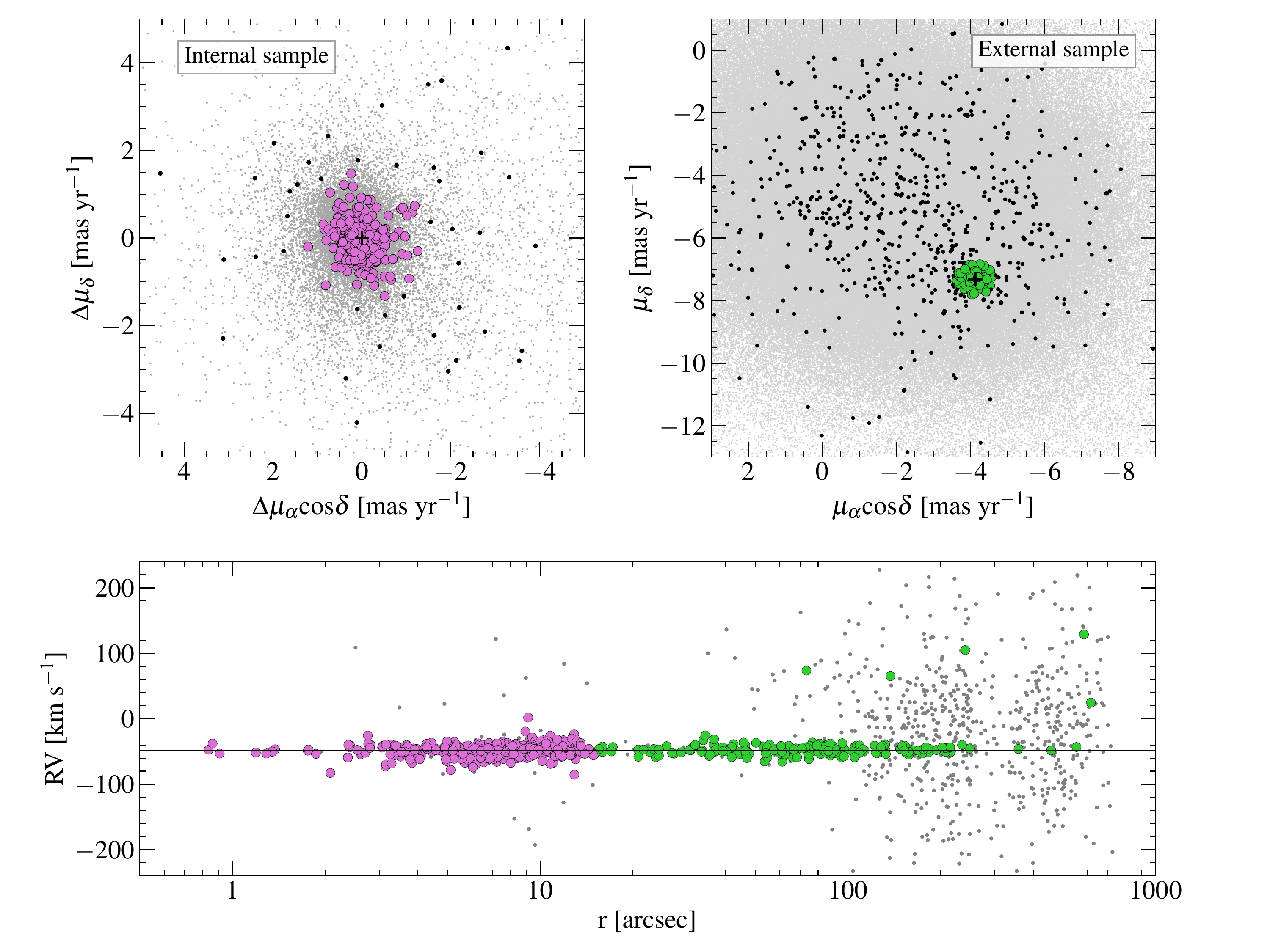}
\centering
\caption{\textit{Top-left panel}: VPD of the relative PMs obtained in
  \citet[][gray dots]{saracino+19}.  The member stars selected from
  the internal sample for the kinematic anaklysis of the cluster are
  marked with large  magenta circles, while the black dots are the targets
  rejected as non-members. The black cross is centered on (0,0), thus
  marking the bulk motion of the cluster. {\it Top-right panel}: VPD
  of the Gaia EDR3 dataset (gray dots,  only stars with $g<18$ are plotted for visualization purposes). The green circles show the
  targets of the external sample selected as member stars, while those
  considered as field stars are indicated with black dots.  The black
  cross marks the absolute PM of NGC 6569 \citep{vasiliev+21}.  {\it
    Bottom panel}: RVs of the 1292 targets of the final catalog as a
  function of the distance from the cluster center.  The large circles
  mark the targets selected as cluster members on the basis of the
  measured PMs, color-coded as in the top panels, while the gray dots
  are the targets rejected as field stars.  }
\label{fig:members}
\end{figure}

In order to properly study the internal kinematics of the cluster, we
have performed an accurate selection of member stars among the targets
of our final catalog.  To obtain the most reliable cluster membership
selection, we used a combination of PMs provided by the Gaia EDR3
\citep{gaia2016,gaiaEDR3} and the relative PMs measured by \citet{saracino+19}.
In particular, for the external sample (FLAMES and KMOS), we adopted
the Gaia PMs by selecting as cluster members the targets with
magnitude $G<19$ and PMs within 0.5 mas yr$^{-1}$ from the absolute PM
of the cluster \citep{vasiliev+21} in the vector-point diagram
(VPD). The Gaia VPD and the member selection of the external sample
are shown in the top-right panel of Figure \ref{fig:members}.  Most of
the targets of the inner sample (MUSE and SINFONI), instead, are
located in regions too crowded to allow reliable Gaia PM measures. We
thus used the relative PMs obtained by \citet{saracino+19} from the
combination of HST and GeMS observations secured at two different
epochs.
In selecting member stars for this sample we followed the
prescriptions described in \cite{saracino+19}: the targets of the
internal sample selected as member stars are shown with large magenta
circles in the VPD of the relative PMs in the top-left panel of Figure
\ref{fig:members}.
The efficiency of the adopted PM selection in excluding field stars is
evident from the bottom panel of Figure \ref{fig:members}, which shows
the target RVs as a function of the distance from the center, with the
stars selected as cluster members marked with colored large circles.
After this selection, the bulk of cluster members is centered at about
$-48$ km s$^{-1}$, while the number of contaminating field stars (gray
dots in the figure) is significant at larger radii and becomes
dominant in the outermost regions.  The remaining obvious outliers
have been removed in the following analysis, according to their
RV. With the purpose to perform a precise kinematic analysis, we also
adopted additional criteria to select the targets with the most
reliable RV measurements\footnote{The RVs sample with the
corresponding errors is available for free download at:
\url{http://www.cosmic-lab.eu/Cosmic-Lab/MIKiS_Survey.html},   see Table \ref{tab:RV} for a preview.}.
Therefore, for the following analysis, we used only the cluster member
targets with S/N$>15$ and RV error $<5$ km s$^{-1}$ (see their
position on the plane of the sky and in the CMDs in
Figs. \ref{fig:mappa} and \ref{fig:cmd}, respectively).

\subsection{Systemic velocity}
\label{sec:vsys}
\begin{figure}[ht!]
\centering
\includegraphics[width=0.9\textwidth]{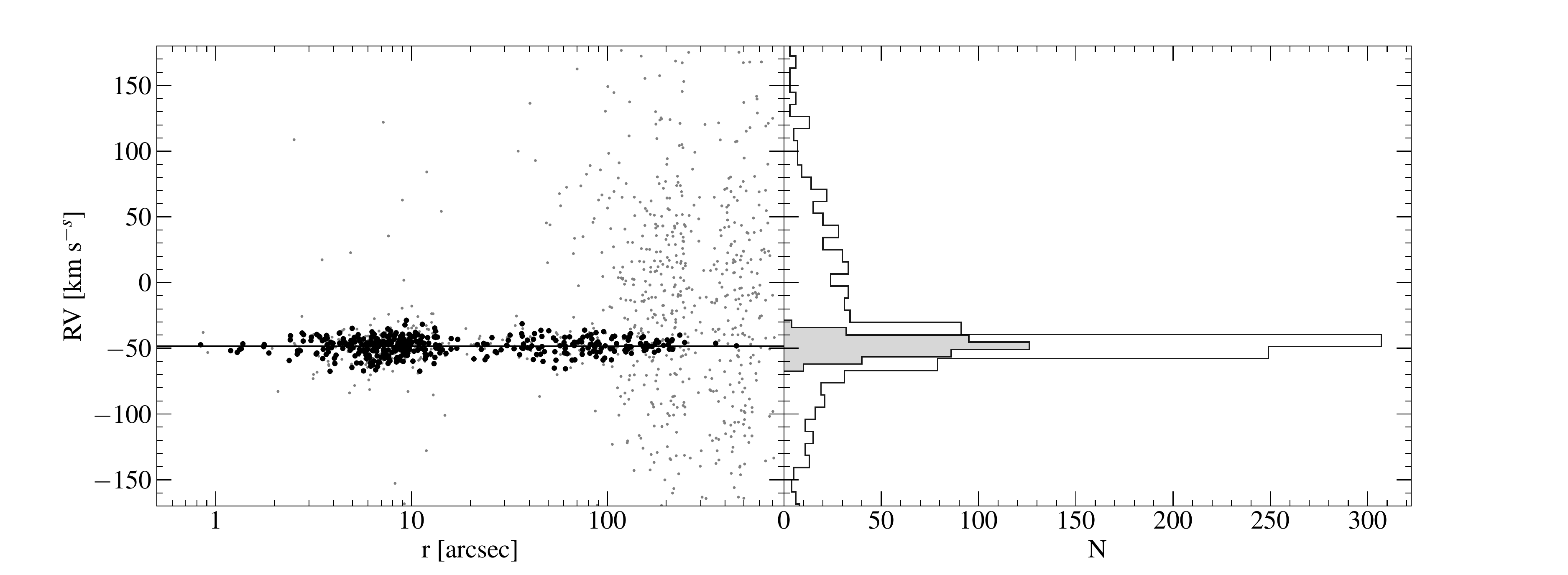}
\centering
\caption{{\it Left panel:} RVs of the final catalog as a function of
  the distance from the cluster center.  The black circles show the
  sub-sample of stars used for the determination of $V_{\rm sys}$
  (solid line), while the gray dots mark the excluded targets.  {\it
    Right panel:} the empty histogram is the number distribution of
  the entire RV sample, while the gray histogram refers to the sample
  used to determine $V_{\rm sys}$ (black dots in the left panel).}
\label{fig:rv_dist}
\end{figure}

For the determination of the systemic velocity ($V_{\rm sys}$) of NGC
6569, among the targets selected with the criteria described above, we
conservatively adopted additional cuts in RVs ($-80$ km
s$^{-1}<$RV$<-20$ km s$^{-1}$) and applied a $3\sigma$-clipping
algorithm to the remaining distribution, thus minimizing the risk of
residual field contamination.  The RVs of the resulting sample (made
of 393 stars) are plotted as black circles as a function of the
distance from the center in the left panel of Figure
\ref{fig:rv_dist}, while their distribution is drawn as fill gray
histogram in the right panel, with the peak indicating the systemic
velocity of the cluster.
Assuming a Gaussian RV distribution, we determined the value of
$V_{\rm \rm sys}$ and its uncertainty by means of a Maximum-Likelihood
procedure \citep{Walker+06}, obtaining $V_{\rm \rm sys} = -48.5 \pm
0.3$ km s$^{-1}$.    Our estimate is in agreement with the value
derived in \citet[][$-47 \pm 4$ km s$^{-1}$]{valenti+11} and,
marginally, with that obtained by \citet[][$-49.9 \pm 0.4$ km
  s$^{-1}$]{Baumgardt+18}, while it strongly disagrees with that
listed in \citet[][$-28.1 \pm 5.6$ km s$^{-1}$]{harris+96}.

\subsection{Systemic rotation}
\label{sec:rot}
In previous kinematics analyses no clear signals of rotation have been
detected in the external regions of NGC 6569
\citep[e.g.,][]{sollima+19, vasiliev+21}.  However, the large sample
of RVs data presented here offers the opportunity to push the search
for rotation further, also including the central region of the
cluster. To this purpose, we used the method already adopted in
several papers by our group \citet{ferraro+18b, lanzoni+18a,
  lanzoni+18b, leanza+22, leanza+22b}.

The method is fully described in \citet[][see also
  \citealp{lanzoni+13}]{bellazzini+12} and consists in splitting the
observed RV sample into two sub-samples on either side of a line
passing through the cluster center. The position angle (PA) of the
line is then changed from $0\arcdeg$ (North direction) to $180\arcdeg$
(South direction), by steps of $10\arcdeg$ and with PA $=90\arcdeg$
corresponding to the East direction.  If the cluster is not rotating,
the mean velocity of the two sub-samples is the same, while in the
presence of systemic rotation along the LOS, it is maximum for the
value of PA corresponding to the rotation axis (PA$_0$). Hence, a
coherent sinusoidal variation of the difference between the mean
velocity of the two sub-samples ($\Delta V_{\rm mean}$) as a function
of PA can be used as diagnostic of rotation, with the maximum/minimum
value of this curve providing twice the rotation amplitude ($A_{\rm
  rot}$) and the position angle of the rotation axis (PA$_0$). In
addition, the distribution in a diagram showing the stellar RV as a
function of the projected distances from the rotation axis (XR) is
expected to appear highly asymmetric, with two diagonally opposite
quadrants being more populated than the other two. Finally, the
cumulative RV distributions of the two sub-samples of stars on each
side of the rotation axis are expected to be different, and the
statistical significance of such difference can be evaluated through
various estimators.  Here we adopt the following three: the
probability that the RV distributions of the two sub-samples are
extracted from the same parent family is estimated by means the
p-value of the Kolmogorov-Smirnov (KS) test, while the statistical
significance of the difference between the two sample means is
evaluated with both the Student's t-test and a Maximum-Likelihood
approach.

Of course a meaningful application of this method requires a uniform
distribution of the RV measures in the plane of the sky.  Thus, we are
forced to avoid some heavy under-sampled regions and limit the
analysis to the innermost $5\arcsec$, where the combination of the
MUSE and SINFONI targets offers a reasonably symmetric coverage (see
Figure \ref{fig:mappa}), and the annular region between 15 and
$150\arcsec$, which is sampled by FLAMES and KMOS data.

\begin{figure}[ht!]
\centering
\includegraphics[width=1.02\textwidth]{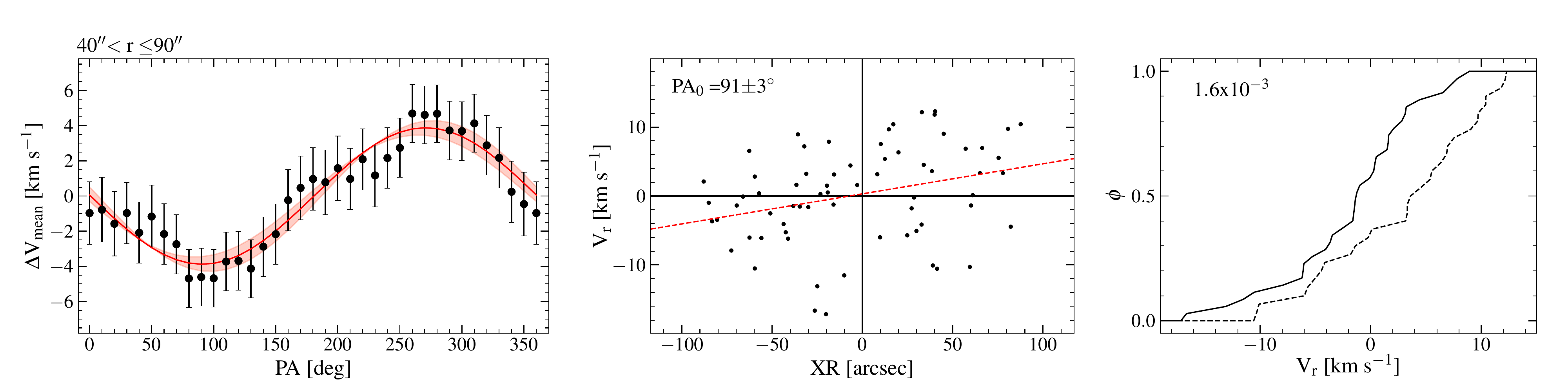}
\centering
\caption{Diagnostic diagrams of the most significant rotation signal
  detected in NGC 6569, in the annular region $40\arcsec<r<90\arcsec$
  from the cluster center.  The {\it left panel } shows the difference
  between the mean RV on each side of a line passing through the
  center with a given PA, as a function of PA itself. The continuous
  line is the sine function that best fits the observed patterns.  The
  {\it central panel} shows the distribution of the velocities
  referred to $V_{\rm sys}$ ($V_r$), as a function of the projected
  distances from the rotation axis (XR) in arcseconds.  The value of
  PA$_0$ is labeled.  The red dashed line is the least square fit to
  the data.  The {\it right panel} shows the cumulative $V_r$
  distributions for the stars with XR$<0$ (solid line) and for those
  with XR$>0$ (dotted line). The Kolmogorov-Smirnov probability that
  the two sub-samples are extracted from the same parent distribution
  is labelled.}
\label{fig:rot}
\end{figure}

The maximum signal of rotation in these regions has been detected at
$40\arcsec<r<90\arcsec$.  This is shown in the diagnostic diagrams
plotted in Figure \ref{fig:rot}, which indicate a maximum amplitude of
$\sim (1.9\pm0.3)$ km s$^{-1}$, a position angle of the rotation axis
PA$_0= (91\pm3)\arcdeg$, and a p-value of the KS test p=0.0016,
indicating that the difference between cumulative RV distributions of
the two
sub-samples on either side of the rotation axis is significant at $\sim 2.4\sigma$.
However, the number of stars observed in this region is admittedly
small (67), and additional spectroscopic observations are needed to
solidly confirm the presence of systemic rotation in this GC.

\subsection{Velocity dispersion profile}
\label{sec:vdp}
As discussed in previous papers \citep[e.g.,][]{lanzoni+18a,
  lanzoni+18b, leanza+22}, the measure of the observed RV dispersion at
different radial distances from the center corresponds to the second
velocity moment profile $\sigma_{II}(r)$, which is linked to the
velocity dispersion profile $\sigma_P(r)$ through the following
relation: $\sigma^2_P(r) = \sigma^2_{II}(r) - A^2_{\rm rot}(r)$.
Since the evidence of rotation in NGC 6569 is highly uncertain (see
previous Section), we assume that the rotation contribution is
negligible and $\sigma^2_P(r) = \sigma^2_{II}(r)$.

\begin{figure}[ht!]
\centering
\includegraphics[width=0.7\textwidth]{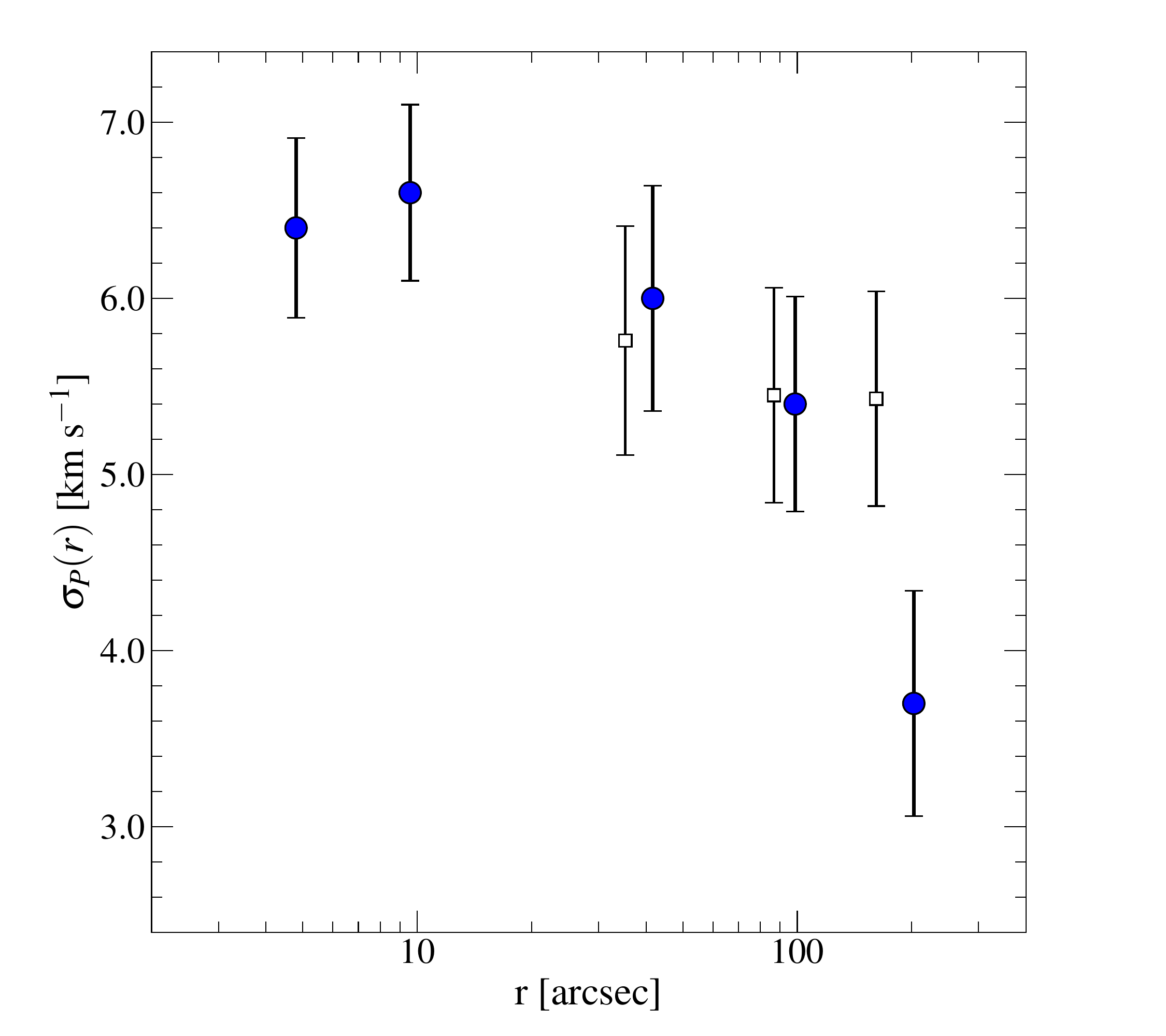}
\centering
\caption{Velocity dispersion (second velocity moment) profile of NGC
  6569 (blue circles) obtained from the measured individual RVs.  For
  the sake of comparison, the white squares correspond to the profile
  published in \citet{Baumgardt+18}.
}
\label{fig:vdp}
\end{figure}

Hence, to determine the velocity dispersion profile of the cluster, we
used the sample of RVs selected with the criteria described in Section
\ref{sec:vsys} (only member stars, with S/N$>15$ and RV error $<5$
km s$^{-1}$).  This has been divided into concentric radial bins with
increasing distance from the center, ensuring both a proper radial
sampling and a statistically sufficient number of targets { (at least
25)} in each bin.  After a 3$\sigma-$clipping procedure used to exclude
the obvious outliers, we then determined the velocity dispersion value
in each bin by following the Maximum-Likelihood method described in
\citet[][see also \citealp{Martin_2007};
  \citealt{sollima+09}]{Walker+06}.  The velocity dispersion
uncertainties are estimated adopting the procedure described in
\citet[][]{pryor+93}. The results are shown in Figure \ref{fig:vdp}
(blue circles) and listed in also Table \ref{tab:vdp}.
As apparent, the velocity dispersion decreases from a central value of
approximately 6.5 km s$^{-1}$, down to 3.7 km s$^{-1}$ in the bin
centered at $\sim 200\arcsec$.  For the sake of comparison, in the
figure we also report the profile published\footnote{ Note that the comparison has been done with published values, however the online repository is repeatedly revised and some updated values are in better agreement with our determinations.} in \citet[][white
  squares]{Baumgardt+18}.  Their two innermost points are consistent with our measures, but they are limited at $r>30\arcsec$ and therefore do not properly characterize the central
 portion of the velocity dispersion profile. The outermost measure from these authors
largely exceeds the dispersion velocity obtained here, probably
because of residual field contamination.

\begin{deluxetable*}{RRRRCC}
\tablecaption{Velocity dispersion profile of NGC 6569.}
\tablewidth{0pt}
\tablehead{
\colhead{ $r_i$ } & \colhead{ $r_e$ }  & \colhead{$r_m$}  &
\colhead{$N$} & \colhead{$\sigma_P$} & \colhead{$\epsilon_{\sigma_P}$} \\  
\colhead{ [arcsec] } & \colhead{ [arcsec] }  & \colhead{[arcsec]}  &  \colhead{ } &\colhead{km s$^{-1}$ }  &
\colhead{ km s$^{-1}$} 
}
\startdata
0.01  &  7.00  &  4.80  & 122 & 6.40 &  0.51  \\
7.00  &  15.00 &  9.58  & 131 & 6.60 &  0.50  \\
15.00 &  70.00 & 41.56  &  65 & 6.00 &  0.64  \\
70.00 & 150.00 & 98.72  &  50 & 5.40 &  0.61 \\
150.00 &550.00 & 202.63 & 26  & 3.70 &  0.64  \\
\enddata
\tablecomments{The table lists: the internal and external radius of
  each radial bin ($r_i$ and $r_e$, respectively), the average
  distance from the center of the stars within the bin ($r_m$), the
  number of stars in the bin ($N$), the measured velocity dispersion
  and its uncertainty in the bin ($\sigma_P$ and
  $\epsilon_{\sigma_P}$, respectively).}
\label{tab:vdp}
\end{deluxetable*}

\begin{figure}[ht!]
\centering
\includegraphics[width=0.7\textwidth]{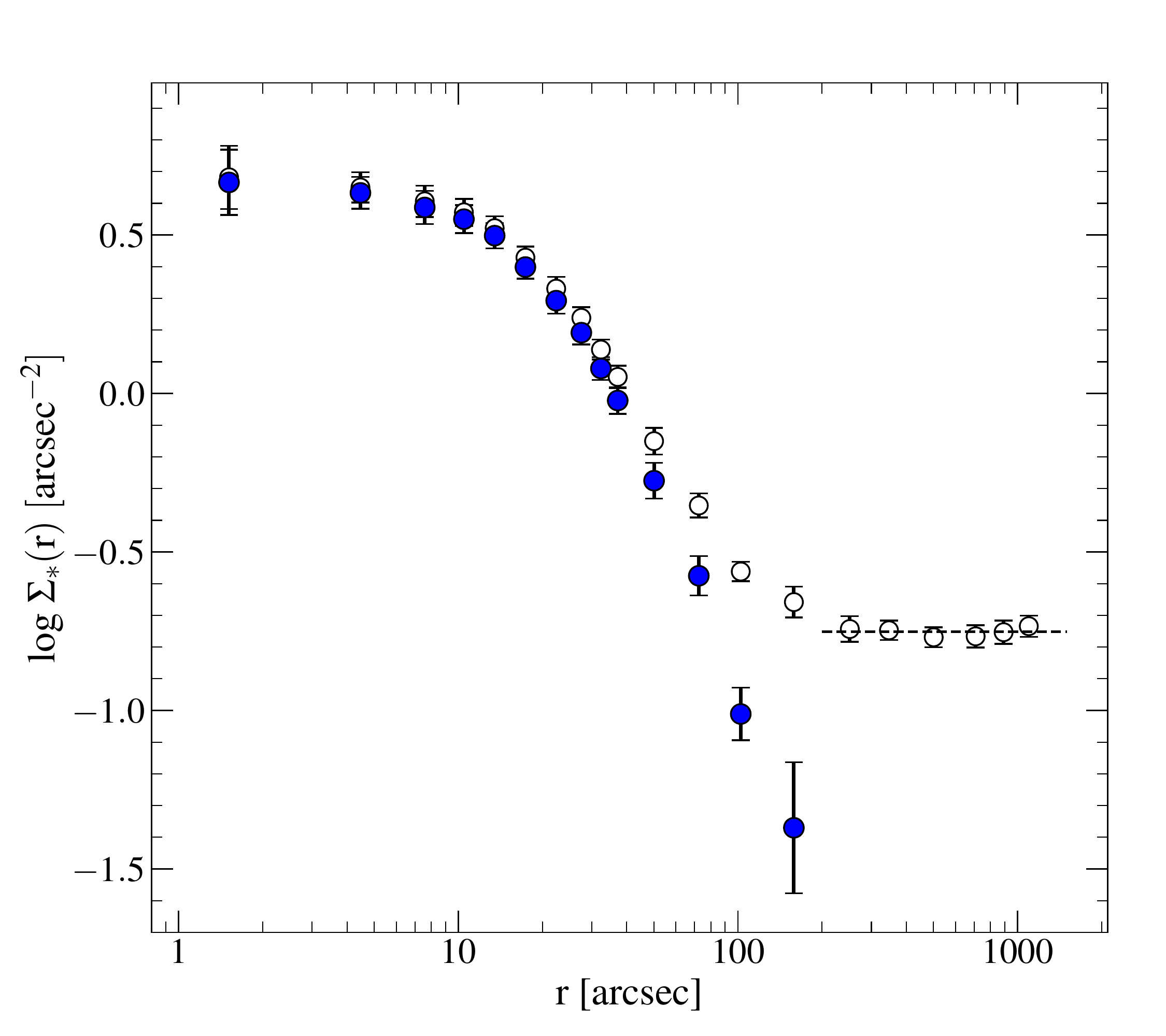}
\centering
\caption{Observed density proﬁle of NGC 6569 obtained from resolved
  star counts (empty circles). The 6 outermost points (with
  $r>200\arcsec$) define a sort of plateau and have been used to
  measure the density level of the background (dashed line). The blue
  circles show the cluster density profile obtained after subtraction
  of the ﬁeld contribution.}
\label{fig:densp}
\end{figure} 
\begin{figure}[ht]
\centering
\includegraphics[width=0.99\textwidth]{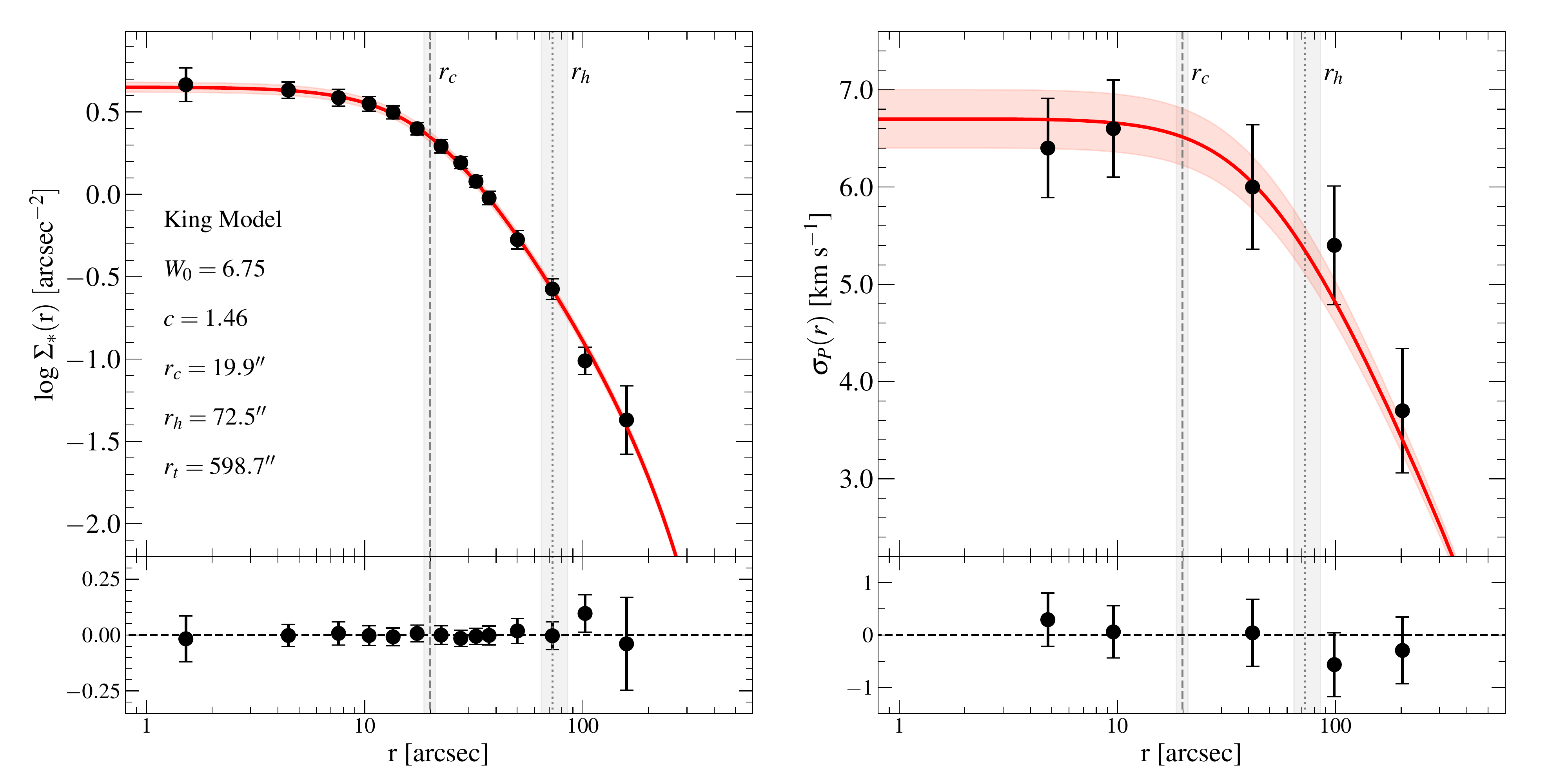}
\centering
\caption{Best-fit King model (as derived in Section \ref{sec:discussion};
  red solid lines) overplotted to the star density proﬁle (left panel)
  and the velocity dispersion profile (right panel) of NGC 6569. The
  best-fit values of the central dimensionless potential ($W_0$),
  concentration parameter ($c$), and core, half-mass and tidal radii
  ($r_c$, $r_h$ and $r_t$, respectively) are labeled in the left
  panel, where the dashed and dotted lines mark $r_c$ and $r_h$ ,
  respectively.  The shaded regions show the associated uncertainties.
  The bottom panels show the residuals between the model and the
  observations. The red shaded region in the right panel marks the
  1$\sigma$ confidence level on the estimate of the central velocity
  dispersion (see Section \ref{sec:discussion}).}
\label{fig:king}
\end{figure} 

\section{Star density profile}
\label{sec:density}
The HST data-set described in \citet{saracino+19} offers the ideal
 angular resolution for determining the innermost portion
($r<120\arcsec$) of the projected density profile of NGC 6569 from
resolved star counts. To cover the entire extension of the system, we
complemented this dataset with the Gaia EDR3 catalog out to
$1200\arcsec$ from the cluster center.  To ensure comparable levels of
photometric completeness  in the two datasets, and sufficiently large
statistics in each bin, we adopted two different magnitude cuts for
the sample selection:
$\rm m_{F814W}<20.5$ for the HST/GeMS dataset, and $G<18$ for the Gaia one.

According to the standard procedure used in several previous works
\citep[see, e.g.][]{lanzoni+10, lanzoni+19, miocchi+13, pallanca+21},
we divided the sample into 20 concentric annuli centred on the
$C_{\mathrm{grav}}$ and split each annulus into an appropriate number
of sectors (usually four).  We estimated the star surface density in
each sector as the ratio between the number of stars within the sector
and the area of the sector itself. The stellar density in each annulus
is then obtained as the average of the sectors densities, while their
standard deviation is adopted as corresponding uncertainty.  Owing to
the different magnitude cuts adopted for the sample selections, the
external portion of the profile (from Gaia data) has been finally
shifted until it matched the last point of the HST/GeMS profile.  The
projected density profile thus obtained is shown in Figure
\ref{fig:densp} (open circles), where the radius associated to each
annulus is the midpoint of the radial bin.  The plateau at
$r>200\arcsec$ is due to the (dominant) contribution of the Galactic
ﬁeld, which has an essentially constant density at the small scales
surveyed here. Its value has been estimated as the averaging density
of the 6 outermost points (dashed line in Figure \ref{fig:densp}), and
it has been subtracted from the observed profile. The results are
shown as blue circles in Figure \ref{fig:densp}.  As apparent, after
the background subtraction the outermost portion of the profile
significantly decreases with respect to the observed one, which
demonstrates as an accurate determination of the field level is
essential to properly constrain the true density distribution of the
system. This has the typical shape observed for most GCs, i.e., a
inner flat core, followed by a steady decreasing trend (Figure
\ref{fig:densp}).

\section{Discussion and Conclusions}
\label{sec:discussion}
The star density and velocity dispersion profiles presented in
Sections \ref{sec:density} and \ref{sec:vdp}, respectively, have been
obtained by using samples of stars with approximately the same
mass. In addition, although no definitive conclusions on the possible
existence of systemic rotation can be drawn at the moment, the results
discussed in Section \ref{sec:rot} suggest that the rotational
velocity (if any) is negligible in this system. Hence, we determined
the structural and kinematic parameters of NGC 6569 by simultaneously
fitting the density and velocity dispersion profiles with single-mass,
spherical, isotropic and non-rotating King models \citep{king+96}.
These are a mono-parametric family of dynamical models, univocally
characterized by the value of the dimensionless parameter $W_0$, which
is proportional to the gravitational potential at the center of the
system, or, equivalently, by the concentration parameter $c$, deﬁned
as $c\equiv\mathrm{log}(r_t/r_0)$, where $r_t$ and $r_0$ are the tidal
and the King radii of the model, respectively.  To determine the
best-fit King model to both the profiles we adopted a Markov Chain
Monte Carlo (MCMC) approach, by means of the emcee algorithm
\citep{Foreman+13}.  We assumed uniform priors on the parameters of
the fit, obtaining the posterior probability distribution functions
(PDFs) for the characteristic parameters of the King models and for
the central velocity dispersion, $\sigma_0$.  For each parameter, the
PDF median has been adopted as best-fit value, while the 16-th and
84-th percentiles of the posterior PDF have been used to estimate the
$1\sigma$ uncertainty.

The resulting best-fit King model is shown overplotted to the observed
density and velocity dispersion profiles in Figure \ref{fig:king} (red
lines), where the residuals between the model and the observations are
also shown in the bottom panels. A very good agreement is apparent,
thus indicating that both the structure (as it is often the case for
Galactic GCs) and the internal kinematics of NGC 6569 are well
consistent with the King model expectations.  The best-fit model is
characterized by $W_0 = 6.75$ (which corresponds to $c=1.46$) and a
core radius $r_c=19.9\arcsec$, corresponding to $\sim 1$ pc at the
distance of the cluster ($d=10.1$ kpc; \citealp{saracino+19}).  The
half-mass and tidal radii are, respectively, $r_h=72.5\arcsec$ and
$r_t=589.7\arcsec$, while the effective radius (i.e., the radial
distance at which the projected density, or the surface brightness,
halves the central value) is $r_{\mathrm{eff}}=54.1\arcsec$. For the
central velocity dispersion we obtained $\sigma_0 = 6.7 \pm 0.3$ km s
$^{-1}$. The best-fit values and the uncertainties of each parameter
are listed in Table \ref{tab_final}.
 
Comparing our estimates with the results obtained from the surface
brightness proﬁle by \citet[][which are also the values quoted in
  \citealt{harris+96}]{Mclaughlin+05}, we find consistent values
within the uncertainties: in fact, after conversion from parsec to
arcseconds using the cluster distance provided in that paper ($d =
10.7$ kpc), these authors quote: $W_0 = 6.20 \pm 0.2$, $c=1.31 \pm
0.05$, $r_c=21 \arcsec$, $r_{\mathrm{eff}}=48\arcsec$, and
$r_t=461\arcsec$.  Within the errors, our values are in agreement also
with those of \citet[][$r_c=20.74\arcsec$ and $r_h=78.86\arcsec$,
  converted into arcseconds using the cluster distance quoted in their
  paper: $d = 12.0$ kpc]{Baumgardt+18}.  We remark, however, that this
comparison is not obvious, nor rigorous, because \citet{Baumgardt+18}
derive these quantities from $N$-body simulations, instead of fitting
the observations with King models.  As for the estimate of $\sigma_0$,
our value is lower than that quoted in \citet[][$\sigma_0\sim7.5$ km
  s$^{-1}$]{Baumgardt+18}.  This discrepancy is likely due to the fact
that the velocity dispersion profile used in that work is poorly
constrained (see Figure \ref{fig:vdp}, and also Figure E12 in
\citealt{Baumgardt+18}).

Under the (well motivated) assumption of a King model structure and by adopting the obtained value of $\sigma_0$, we estimated the total mass of the cluster 
as  $M= 166.5 r_0 \mu/\beta$ \citep{Majewski+03}, where $r_0$ is the King radius, $\mu$ is a scaling parameter depending on the King concentration $c$ as
$\log \mu = -0.14192 c^4 +1.15592 c^3 -3.16183 c^2 +4.21004 c -1.00951$ \citep{Djorgovski+93}, and $\beta=1/\sigma_0^2$ \citep{Richstone+86}.   
The uncertainty has been estimated as the dispersion of the mass values
resulting from 1000 Monte Carlo simulations, run by adopting
appropriate normal distributions  for $c$, $r_0$ and $\sigma_0$
\citep[see][]{leanza+22}.  The resulting total mass 
is $M =1.72_{-0.18}^{+0.20} \times 10^5 M_\odot$. This is lower than the
value obtained by \citet[][$3.02 \pm 0.36 \times 10 ^5$
  M$_\odot$]{Baumgardt+18} from $N$-body simulations, presumably due
to the different velocity dispersion profile used as constraint, other
than the different assumptions and methods adopted in the two works
(for instance, \citealp{Baumgardt+18} adopt a 20\% larger distance of
the cluster than assumed here, and they compare the observations with
a library of N-body simulations).
Finally, by adopting the structural parameters determined here, the total mass and the cluster distance of \citet{saracino+19}, we estimated the central and half-mass relaxation times ($t_{\rm rc}$ and $t_{\rm rh}$) from equation (10) and equation(11) of \citet{Djorgovski+93}, respectively. To estimate the central relaxation time (in years), we used:
\begin{equation}
t_{\rm rc} = \frac{0.834 \times10^7}{\ln(0.4 N)} \frac{\rho_{0,M}^{1/2} r_c^3}{m},
\end{equation}
where $N$ is the total number of stars, $\rho_{0,M}$ is the central mass density, and $m$ is the average stellar mass.
For the half-mass relaxation time (in years), we adopted:
\begin{equation}
t_{\rm rh} = \frac{2.055 \times10^6}{\ln(0.4 N)} \frac{M^{1/2} r_h^3}{m}
\end{equation}
We found 
$\log(t_{\rm rc})=8.3$ and   $\log(t_{\rm
  rh})=9.2$, which are roughly  consistent with the
values quoted in the \citet[][8.38 and 9.05, respectively]{harris+96}
catalog, consistently with the fact that also the differences in terms
of structural parameters and cluster distance are small. These values
suggest that NGC 6569 is in an intermediate stage of its dynamical
evolution, although dedicated investigations of this issue (see, e.g.,
\citealp{ferraro+18c} and \citealp{bhat+22, bhat+23,cadelano+20_m15m30}) are needed to
properly confirm it.

\begin{deluxetable*}{lll}
\tablecaption{Summary of the main parameters used and obtained in this work for the GGC NGC 6569. \label{tab_final}}
\tablewidth{0pt}
\renewcommand{\arraystretch}{1.2}
\tablehead{
\colhead{Parameter} & \colhead{Estimated Value} & \colhead{Reference}
}
\startdata
    Cluster center & $\alpha_{\mathrm{J2000}} = 18^{\mathrm{h}} 13^{\mathrm{m}} 38.70^{\mathrm{s}} $ & this work\\ [-5pt]
                        & $\delta_{\mathrm{J2000}} = -31\arcdeg 49\arcmin 37.13\arcsec $ \\
    Metallicity & [Fe/H]$=-0.79$ &  \citet{valenti+11} \\
                     & [$\alpha$/Fe]$=+0.4$ &  \citet{valenti+11} \\ 
    Cluster distance & $d=10.1 \pm 0.2$ kpc &  \citet{saracino+19}\\
    Dimensionless central potential  & $W_0 = 6.75\pm 0.40$ & this work  \\
    Concentration parameter & $c=1.46_{-0.11}^{+0.12}$ & this work  \\
    Core radius & $r_c=19.9\arcsec \pm 1.2 = 0.97$ pc & this work  \\
    Three-dimensional half-mass radius & $r_{\rm h}=72.5\arcsec_{-7.9}^{+12.9}=3.55$ pc& this work  \\
    Effective radius & $r_{\mathrm{eff}}=54.1\arcsec_{-5.7}^{+9.3} = 2.65$ pc & this work \\
    Tidal radius & $r_t=589.7\arcsec_{-109.3}^{+167.9} = 28.88$ pc &  this work \\
    Systemic velocity & $V_{\rm sys} = -48.5 \pm 0.3$ km s$^{-1}$ & this work \\
    Central velocity dispersion & $\sigma_0 = 6.7 \pm 0.3$ km s$^{-1}$ & this work  \\
    Absolute proper motions & $\mu_\alpha cos\delta =-4.125$ mas yr$^{-1}$  &   \citet{vasiliev+21} \\ [-5pt]
                         & $\mu_\delta= -7.315$ mas yr$^{-1}$& \\
    Total mass & $M = 1.72_{-0.18}^{+0.20} \times 10^5 M_\odot$ &  this work \\
    Central relaxation time & $\log(t_{\rm rc})=8.3$ [in yr] &  this work \\
    Half-mass relaxation time & $\log(t_{\rm rh})=9.2$ [in yr] &  this work \\
\enddata
\end{deluxetable*}

\newpage
\appendix{}
\section{Kinematic catalog}

Table \ref{tab:RV} provides the kinematic catalog of the quality selected stars  (S/N$>15$ and RV error $<5$ km s$^{-1}$) in NGC 6569. 

\begin{deluxetable*}{RCCCCCCCC}[h]
\tablecaption{Kinematic catalog}
\tablewidth{0pt}
\tablehead{
\colhead{ id } & \colhead{ $\alpha$ }  & \colhead{$\delta$}  &
\colhead{$\rm m_{F555W}$} & \colhead{$\rm m_{F814W}$} & \colhead{J}  & \colhead{K} & \colhead{RV} & \colhead{$\rm \epsilon_{RV}$} \\  
\colhead{ } & \colhead{ [deg] } & \colhead{ [deg] }  & \colhead{}  &  \colhead{ } & \colhead{}  &  \colhead{} &\colhead{[km s$^{-1}]$ }  &
\colhead{[km s$^{-1}]$} 
}
\startdata
      1  &   273.4090654  &   -31.8265292  &   15.17  &   12.64  &   11.23  &   10.07  &    -60.6  &	 0.3   \\ 
     2  &   273.4110241  &   -31.8276114  &   15.18  &   12.69  &   11.26  &   10.11  &    -59.3  &	 0.3   \\ 
     3  &   273.4094891  &   -31.8281725  &   15.25  &   12.76  &   11.36  &   10.25  &    -47.1  &	 0.3   \\ 
     4  &   273.4088372  &   -31.8290447  &   15.22  &   13.01  &   11.57  &   10.41  &    -37.9  &	 0.8   \\ 
     5  &   273.4109841  &   -31.8251144  &   15.53  &   13.43  &   12.12  &   11.15  &    -44.9  &	 0.7   \\ 
     6  &   273.4112084  &   -31.8295540  &   16.01  &   14.03  &   12.81  &   11.92  &    -44.6  &	 0.8   \\ 
     7  &   273.4121423  &   -31.8295264  &   15.86  &   13.96  &   12.90  &   12.00  &    -46.2  &	 0.7   \\ 
     8  &   273.4080336  &   -31.8266938  &   16.29  &   14.31  &   13.01  &   12.07  &    -47.2  &	 1.1   \\ 
     9  &   273.4117724  &   -31.8302748  &   16.22  &   14.23  &   13.02  &   12.11  &     84.2  &	 0.9   \\ 
    10  &   273.4098361  &   -31.8293371  &   16.49  &   14.52  &   13.18  &   12.29  &    -55.5  &	 1.0   \\ 
    ...  &  ... &  ... &  ...  &  ...  & ... &  ...  &    ...  &    ...  \\
\enddata
\tablecomments{ First 10 rows of the kinematic catalog, reporting  the identification number, the  absolute coordinates, the optical and near-infrared magnitudes (if available), and the measured line-of-sight velocities and errors for all the stars surviving the quality selection criteria. No membership selection is applied. The entire table is provided in the online version.}
\label{tab:RV}
\end{deluxetable*}

\vskip1truecm We thank the anonymous referee for the useful comments that improved the paper.
This work is part of the project {\it Cosmic-Lab} at the
Physics and Astronomy Department “A. Righi” of the Bologna University
(http://www.cosmic-lab.eu/ Cosmic-Lab/Home.html). The research was
funded by the MIUR throughout the PRIN-2017 grant awarded to the
project {\it Light-on-Dark} (PI:Ferraro) through contract
PRIN-2017K7REXT. 
This work has made use of data from the European Space Agency (ESA) mission
{\it Gaia} (\url{https://www.cosmos.esa.int/gaia}), processed by the {\it Gaia}
Data Processing and Analysis Consortium (DPAC,
\url{https://www.cosmos.esa.int/web/gaia/dpac/consortium}). Funding for the DPAC
has been provided by national institutions, in particular the institutions
participating in the {\it Gaia} Multilateral Agreement.



\bibliographystyle{aasjournal}


\end{document}